\documentclass[twocolumn,preprintnumbers,superscriptaddress,endnote,nofootinbib,aps,prd]{revtex4-1}
\pdfoutput=1

\usepackage{graphicx}
\usepackage{dcolumn}   % needed for some tables 
\usepackage{amssymb}   % for math
\usepackage{amsmath}
\usepackage[FIGTOPCAP]{subfigure}
\usepackage{float}

\usepackage{colordvi}
\usepackage{epsfig}
\usepackage{latexsym}

\newcommand{ \centeron }[2]{{\setbox0=\hbox{#1}\setbox1=\hbox{#2}\ifdim
                             \wd1>\wd0\kern.5\wd1\kern-.5\wd0\fi \copy0
                             \kern-.5\wd0\kern-.5\wd1\copy1\ifdim\wd0>\wd1
                             \kern.5\wd0\kern-.5\wd1\fi}}
\newcommand{ \ltap }{\>\centeron{\raise.35ex\hbox{$<$}}
                     {\lower.65ex\hbox{$\sim$}}\>}
\newcommand{ \gtap }{\>\centeron{\raise.35ex\hbox{$>$}}
                     {\lower.65ex\hbox{$\sim$}}\>}

\newcommand{ \slashchar }[1]{\setbox0=\hbox{$#1$}   % set a box for #1
   \dimen0=\wd0                                     % and get its size
   \setbox1=\hbox{/} \dimen1=\wd1                   % get size of /
   \ifdim\dimen0>\dimen1                            % #1 is bigger
      \rlap{\hbox to \dimen0{\hfil/\hfil}}          % so center / in box
      #1                                            % and print #1
   \else                                            % / is bigger
      \rlap{\hbox to \dimen1{\hfil$#1$\hfil}}       % so center #1
      /                                             % and print /
   \fi}                                             %

\newcommand{\gev}{\text{GeV}}
\newcommand{\tev}{\text{TeV}}

\newcommand{\gsim}{\gtrsim}
\newcommand{\ra}{\rightarrow}
\newcommand{\fb}{\text{fb}}
\newcommand{\mh}{m_h}
\newcommand{\re}{\mathrm{Re}}
\newcommand{\im}{\mathrm{Im}}

\newcommand{\luB}{\lambda_u^B}
\newcommand{\ldB}{\lambda_d^B}
\newcommand{\luW}{\lambda_u^W}
\newcommand{\ldW}{\lambda_d^W}

\newcommand{\muu}{\mu_u}
\newcommand{\mud}{\mu_d}

\newcommand{\hupsq}{ | H_u^+|^2}
\newcommand{\huosq}{ | H_u^0|^2}
\newcommand{\Ruosq}{ |R_u^0|^2}
\newcommand{\Rumsq}{ | R_u^-|^2}
\newcommand{\hdmsq}{ | H_d^-|^2}
\newcommand{\hdosq}{ | H_d^0|^2}
\newcommand{\Rdosq}{ | R_d^0|^2}
\newcommand{\Rdpsq}{ | R_d^+|^2}

\newcommand{\hup}{H_u^+}
\newcommand{\huo}{H_u^0}
\newcommand{\Ruo}{R_u^0}
\newcommand{\Rum}{R_u^-}
\newcommand{\hdm}{H_d^-}
\newcommand{\hdo}{H_d^0}
\newcommand{\Rdo}{R_d^0}
\newcommand{\Rdp}{R_d^+}

\newcommand{\hupc}{H_u^{+*}}
\newcommand{\huoc}{H_u^{0*}}
\newcommand{\Ruoc}{R_u^{0*}}
\newcommand{\Rumc}{R_u^{-*}}
\newcommand{\hdmc}{H_d^{-*}}
\newcommand{\hdoc}{H_d^{0*}}
\newcommand{\Rdoc}{R_d^{0*}}
\newcommand{\Rdpc}{R_d^{+*}}

\newcommand{\nn}{\nonumber}

\begin{document}

\pagestyle{plain}

\preprint{FERMILAB-PUB-12-454-T}

\title{Electroweak Baryogenesis in R-symmetric Supersymmetry}

\author{R. Fok}
\affiliation{Department of Physics and Astronomy, York University, 
             Toronto, ON, Canada, M3J 1P3}

\author{Graham D. Kribs}
\affiliation{Department of Physics, University of Oregon,
             Eugene, OR 97403}

\author{Adam Martin}
\affiliation{Theoretical Physics Department, Fermilab, Batavia, IL 60510}
\affiliation{Department of Physics, University of Notre Dame, Notre Dame, 
             IN 46556\,}~\email{visiting scholar}

\author{Yuhsin Tsai}
\affiliation{Laboratory for Elementary-Particle Physics, Cornell University, 
             Ithaca, NY}

%\date{\today}
%%%%%%%%%%%%%%%%%%%%%%%%%%%%%%%%%%%%%%%%%%%%%%%%%%%%%%%%%%%%%%%%%%%%%%%%%%%%

\begin{abstract}

We demonstrate that electroweak baryogenesis can occur in a 
supersymmetric model with an exact $R$-symmetry.  
The minimal $R$-symmetric supersymmetric model contains
chiral superfields in the adjoint representation, giving
Dirac gaugino masses, and an additional set of 
``$R$-partner'' Higgs superfields, giving $R$-symmetric
$\mu$-terms.  New superpotential couplings between the adjoints and  
the Higgs fields can simultaneously increase the strength of the 
electroweak phase transition and provide additional tree-level contributions
to the lightest Higgs mass.  Notably, no light stop is present
in this framework, and in fact, we require both stops to be
above a few TeV to provide sufficient radiative corrections to
the lightest Higgs mass to bring it up to $125$~GeV\@. 
Large CP-violating phases in the gaugino/higgsino sector allow us 
to match the baryon asymmetry of the Universe with no constraints 
from electric dipole moments due to $R$-symmetry.  
We briefly discuss some of the more interesting phenomenology,
particularly of the of the lightest CP-odd scalar.

\end{abstract}
\maketitle

\section{Introduction}
\label{sec:intro}

The origin of the matter asymmetry is a deep mystery that
remains unsolved.  Conditions that can lead to a dynamical asymmetry 
between baryons and anti-baryons were articulated years ago
by Sakharov \cite{Sakharov:1967dj}:  baryon number violation, 
C and CP violation, and a departure from thermal equilibrium.
All three conditions are satisfied by the standard model
as it passes through the electroweak phase transition.
But, the CP violation is too small \cite{Jarlskog:1985ht},
and the phase transition is not strongly first-order
(e.g., \cite{Anderson:1991zb,Dine:1991ck,Arnold:1992rz,Cohen:1993nk,Quiros:1999jp}).

Weak scale supersymmetry has long been known to potentially 
enhance the strength of the electroweak phase transition and 
provide new sources of CP violation 
\cite{Giudice:1992hh,Espinosa:1993yi,Carena:1996wj}.
In the minimal supersymmetric standard model (MSSM), 
this occurs in the presence of a light stop and a light Higgs boson.
Given the recent LHC results interpreted as the existence of
a Higgs boson at $\mh = 125$~GeV \cite{atlashiggs,cmshiggs},
this region is essentially ruled out \cite{Curtin:2012aa,Carena:2012np}.
Methods to strengthen the first-order phase transition beyond the 
MSSM have been widely discussed
\cite{Pietroni:1992in,Kang:2004pp,Carena:2004ha,Menon:2004wv,Ham:2004nv,Funakubo:2005pu,Huber:2006wf,Shu:2006mm,Profumo:2007wc,Ham:2007wc,Fok:2008yg,Carena:2008rt,Das:2009ue,Kang:2009rd,Ham:2010tr,Cheung:2011wn,Kumar:2011np,Kanemura:2011fy,Cohen:2012zz,Kozaczuk:2012xv}.

In this paper we consider a relatively recent framework for
supersymmetry that incorporates an $R$-symmetry, proposed in 
Ref.~\cite{Kribs:2007ac}.  $R$-symmetric supersymmetry features
Dirac gaugino masses, that have been considered 
long ago \cite{Fayet:1978qc,Polchinski:1982an,Hall:1990hq}
and have inspired more recent model building
\cite{Fox:2002bu,Nelson:2002ca,Chacko:2004mi,Carpenter:2005tz,Antoniadis:2005em,Nomura:2005rj,Antoniadis:2006uj,Kribs:2007ac,Amigo:2008rc,Benakli:2008pg,Blechman:2009if,Carpenter:2010as,Kribs:2010md,Abel:2011dc,Frugiuele:2011mh,Itoyama:2011zi} and phenomenology 
\cite{Hisano:2006mv,Hsieh:2007wq,Blechman:2008gu,Kribs:2008hq,Choi:2008pi,Plehn:2008ae,Harnik:2008uu,Choi:2008ub,Kribs:2009zy,Belanger:2009wf,Benakli:2009mk,Kumar:2009sf,Chun:2009zx,Benakli:2010gi,Fok:2010vk,DeSimone:2010tf,Choi:2010gc,Choi:2010an,Benakli:2011kz,Kumar:2011np,Heikinheimo:2011fk,Fuks:2012im,Kribs:2012gx}.
We show that an $R$-symmetric supersymmetric model can 
simultaneously obtain:  a strong enough first order
phase transition; sufficient CP violation with
no difficulties with electric dipole moment (EDM) bounds; 
and a Higgs mass $\mh \simeq 125$~GeV consistent 
with the LHC observations.  Much of these results rely on 
exploiting the additional superpotential couplings among the 
Higgs fields, their $R$-symmetric partners, and the chiral
adjoint fields.  Kumar and Pont\'on also studied electroweak
baryogenesis in a model with an approximate $R$-symmetry \cite{Kumar:2011np}.
Their approach was to reshuffle the $R$-charges of the fields 
such that $\Phi_B H_u H_d$ operator is allowed, where the fermion singlet 
in $\Phi_B$ is the $R$-partner to the bino.  In our approach, we retain
the original $R$-charges defined by the minimal $R$-symmetric 
supersymmetric standard model \cite{Kribs:2007ac},
utilizing new superpotential couplings among the electroweak adjoints,
the Higgs superfields, and the $R$-partner Higgs superfields.  

Supersoft supersymmetry breaking \cite{Fox:2002bu} shares several
ingredients of the $R$-symmetric model.  One positive feature
is the relative weakness of the all hadronic jets plus missing energy 
search bounds from LHC due to the heavy 
Dirac gluino mass \cite{Kribs:2012gx}.  On the flip-side, however,
the usual $D$-term that determines the tree-level contribution
to the lightest Higgs mass is absent, and no $A$-terms
are generated.  Thus, even with some nontrivial modifications to
the model \cite{Fox:2002bu}, it seems rather difficult to reconcile 
the recent LHC observations of $\mh = 125$~GeV \cite{atlashiggs,cmshiggs} 
with the predictions of the supersoft model.  
In contrast, one of central points of our paper is to show
that there are tree-level (and loop-level) contributions to 
the Higgs mass from the same superpotential couplings that allow
the electroweak phase transition to be strengthened.  
These additional contributions imply $R$-symmetry need not be
broken to generate a large enough lightest Higgs mass.  However, 
we will still need a substantial one-loop contribution from
stops with mass $\simeq 3$~TeV to obtain $\mh \simeq 125$~GeV, 
and so some sacrifice in fine-tuning is inevitable.

\section{The Minimal R-Symmetric Supersymmetric Standard Model}

First we review the field content and new couplings present in
the minimal $R$-symmetric supersymmetric standard model (MRSSM).
In the MRSSM, the gauginos acquire Dirac masses through the 
Lagrangian terms
\begin{equation}
\int d^2\theta \sqrt 2\, 
  \frac{\mathcal{W}'_\alpha \mathcal{W}^\alpha_a \, \Phi^a}{\Lambda_{\rm mess}} 
  + h.c. \, ,
\end{equation} 
where $\mathcal W^{\alpha}_a$ is the field strength superfield 
for one of the SM gauge groups (labelled by $a$, $\alpha$ is a spinor index) 
and $\Phi^a$ is a ``$R$-partner'' chiral superfield 
transforming under the adjoint representation 
of the appropriate gauge group with $R$-charge $R[\Phi^a] = 0$.
Supersymmetry breaking is communicated through $R$-symmetry
preserving spurions that include $\mathcal W'_{\alpha}$ 
which parameterizes a $D$-type spurion, 
$\mathcal W'_{\alpha} = \mathbf D\,\theta_{\alpha}$. 
Expanded into components, the above operator becomes
\begin{eqnarray}
\lefteqn{-\frac{\mathbf D}{\Lambda}( \lambda^a \, \psi_a + h.c. 
                          + \sqrt 2\,D_a \, (A^a + {A^a}^*) ) \; =} 
& & \nonumber \\
& & -M_D \, \left( \lambda^a \, \psi_a + h.c. 
                + 2\,\sqrt{2} \, D_a \, \re(A^a) \right), 
\label{eq:dmass}
\end{eqnarray}
that contains the mass term between the gaugino ($\lambda^a$) 
and its ``$R$-partner'' ($\psi_a$) as well as a coupling of the 
real part of the scalar field within $\Phi^a$ to the $D$-term of 
the corresponding gauge group.

The second term in Eq.~(\ref{eq:dmass}) has two important consequences: 
First, the equation of motion for $\re(A^a)$ sets $D_a \equiv 0$ 
for all three SM gauge groups. The Higgs quartic coupling in the 
MSSM is contained in the $SU(2)$ and $U(1)$ $D$-terms, so eliminating 
these terms will clearly have an impact on the Higgs potential.
Second, while the real parts of $A^a$ 
acquire a mass $\mathcal{O}(M_D)$ from Eq.~(\ref{eq:dmass}), 
$\im(A^a)$ remains massless at this level.

In order to enforce $R$-symmetry on the superpotential, 
the Higgs sector of the MRSSM must be enlarged.
The $\mu$-term of the MSSM is replaced by the $R$-symmetric $\mu$-terms
\begin{equation}
W \supset \mu_u\, H_u\, R_u + \mu_d\, R_d\, H_d \, ,
\label{eq:newmu}
\end{equation}
where $R_{u,d}$ are new, $R$-charge $R[R_{u,d}] = 2$ fields that 
transform as $(\mathbf{1},\mathbf{2})_{\mp 1/2}$ under the 
standard model gauge groups.  
This choice of $R$-partners ensures that electroweak symmetry breaking 
by the Higgs fields $H_{u,d}$ does not spontaneously break $R$-symmetry.
The MRSSM also defines the $R$-charges of the matter fields to be 
$R[Q_i,U_i^c,D^c_i,L_i,E^c_i] = 1$, allowing the usual Yukawa couplings
in the superpotential. 

Given the extra matter content, 
there are new superpotential operators \cite{Kribs:2007ac} 
one can write in the $R$-symmetric theory,
\begin{eqnarray}
W \; & \supset & \; \lambda^u_B\, \Phi_B\, H_u\, R_u 
                  + \lambda^d_B\, \Phi_B\, R_d\, H_d \nonumber \\
    &         &{} + \lambda^u_W \Phi^a_W H_u\, \tau^a R_u 
                  + \lambda^d_W \Phi^a_W R_d\, \tau^a H_d \, .
\label{eq:extrayuk}
\end{eqnarray}
Unlike the $\mu$-terms, which are required to achieve experimentally viable chargino masses, there is no direct phenomenology that dictates that the $\lambda_i$ couplings in Eq.~(\ref{eq:extrayuk}) must be nonzero (being superpotential couplings, they will not be generated radiatively if set to zero initially). However,  these $\lambda_i$ couplings play a vital important role in driving the phase transition to be first order. The importance of the $\lambda_i$ couplings can be seen already from the scalar potential;  the operators in 
Eqs.~(\ref{eq:newmu},\ref{eq:extrayuk}) lead to new trilinear 
and quartic operators involving Higgs fields and the scalars in 
$\Phi_B, \Phi_W, R_u, R_d$. 
\begin{equation}
V \; \supset \; \mu^*_u\, (\lambda^{*B}_u\, A^*_B )|H^0_u|^2 + 
              \mu^*_d\, (\lambda^{*B}_d\, A^*_B)\,|H^0_d|^2 + c.c., 
\label{eq:vex}
\end{equation}
Trilnear scalar interactions involving the Higgs multiplets, 
especially those with large couplings, are well known to impact 
the strength of the electroweak phase transition~\cite{Espinosa:1992hp, Pietroni:1992in,Carena:1996wj,Espinosa:1996qw,Carena:1997gx,Carena:1997ki,Profumo:2007wc,Carena:2008rt,Cohen:2011ap,Espinosa:2011ax,Cohen:2012zz}.

Turning to the supersymmetry breaking parameters of the theory, 
scalar soft masses can arise from an additional source of $F$-term 
supersymmetry breaking.  So long as the supersymmetry breaking 
spurions $X$ have $R$-charge $R[X] = 2$, the $R$-symmetry is 
preserved and no Majorana gaugino masses are 
generated.\footnote{$R$-symmetry is not essential here.
Majorana gaugino masses can be avoided as long as $X$ is not a 
singlet \cite{Kribs:2010md}.}  The soft masses from the 
K\"ahler terms are
\begin{eqnarray}
K & \supset & \int d^4\theta 
     \frac{X^{\dag} X \,Q^{\dag}Q}{\Lambda^2_{\rm mess}}, \\
  &         & Q \in \{ Q_i, U^c_i, D^c_i, L_i, E^c_i, H_{u,d}, R_{u,d}, \Phi^a \} \, .
\nonumber
\end{eqnarray}
In addition, holomorphic soft masses for each $\Phi^a$ are of the form
\begin{eqnarray}
\int d^2\theta \; \frac{\mathcal{W}'_\alpha {\mathcal{W}'}^{\alpha} \Phi^a \Phi^a}{\Lambda_{\rm mess}^2} + h.c. \, .
\end{eqnarray}
We assume the coefficients for the holomorphic soft masses are real.
The full set of soft masses for the scalar components of 
$\Phi_B$ and $\Phi_W$ are given in the Appendix in Eq.~(\ref{eq:softmm}).
Soft-breaking, trilinear scalar couplings between the Higgs and 
squarks or sleptons are forbidden by $R$-symmetry.  
For viable phenomenology, we allow the relative size of the 
supersymmetry breaking contributions to be within roughly one
order of magnitude in mass.  

Throughout this paper we will take the Dirac gaugino masses 
to be large. 
This limit simplifies our calculations and is motivated 
by phenomenology.  Specifically, to avoid conflict with precision electroweak 
observables Ref.~\cite{Kribs:2007ac} found the $SU(2)_w$ gaugino masses 
should be larger than $1$~TeV.  
Such heavy electroweak gauginos decouple from the rest of the theory 
and play little role in the electroweak phase transition. 
The higgsino masses in the MRSSM, on the other hand, 
come from $\mu_u, \mu_d$, which we take to be closer to the
electroweak scale.  

Furthermore, heavy Dirac gauginos, when combined with the MRSSM 
Higgs superpotential structure and lack of $A$-terms, 
lead to \emph{significantly} relaxed flavor constraints. 
As shown in Ref.~\cite{Kribs:2007ac}, low-energy, precision observables 
become insensitive to new sources of flavor or CP in the 
supersymmetric sector.  Electric dipole moments induced by one-loop
contributions involving the gauginos and higgsinos are completely
absent.  This allows $\mathcal{O}(1)$ phases in the MRSSM that 
will be important when we consider CP violation and its role 
in baryogenesis in Sec.~\ref{sec:cpv}.

Having reviewed the MRSSM, its typical spectra and constraints, 
we now investigate the strength of the electroweak phase transition.

\section{Qualitative Features of the Phase Transition in the MRSSM}
\label{sec:EWPTqual}

There are several features of the MRSSM which lead one to suspect 
that the phase transition could be different from more familiar 
(supersymmetric) scenarios. First, as a result of the 
superpotential interaction in Eq.~(\ref{eq:extrayuk}), 
there are extra scalar states coupled to the Higgs boson. 
Extra ``higgsphilic'' scalars are known to (potentially) increase 
the strength of the EW phase transition, with the prototypical example 
being the stop squark. However, unlike the stops of the MSSM, 
these MRSSM states are not colored, and they have limited interactions 
with other SM fields. As a result, these extra scalars can be 
quite light and can interact strongly with the Higgs without 
causing any phenomenological problems.  

The second feature is that the tree-level Higgs potential vanishes 
to leading order in $1/M_D$ where $M_D$ is the Dirac gaugino mass for
the bino and wino.  This can be understood as follows: 
in the limit that all other superpotential or soft-breaking interactions 
involving $\re(A^a)$ are absent or negligible, the $D$-term disappears. 
In the MSSM, the $D$-term is the sole source of 
Higgs quartic interactions at tree-level\footnote{Strictly speaking, 
this assumes only renormalizable superpotential terms are included. 
Higher dimensional operators will change this statement, 
as studied in 
Ref.~\cite{Batra:2008rc,Carena:2009gx,Carena:2010cs,Carena:2011dm}.}; 
$D=0$ means the potential is purely quadratic and tree-level 
symmetry breakdown does not occur. 
In the presence of other interactions involving $Re(A^a)$, 
the $D$-terms are not exactly zero and a (non-trivial) tree-level 
Higgs potential is generated. The dimension of the Higgs operators 
that are generated depends on how $Re(A^a)$ interacts, 
but all operators will be accompanied by coefficients with 
at least one power of the large Dirac gaugino mass, $1/M_D$. 
As a simple example demonstrating 
this mechanism, consider the potential 
\begin{align}
V(a,h) &= m_0^2\,h^2 + (M_D\, a - g\, h^2)^2 + \lambda\mu\, a\,h^2 + m^2_S\, a^2.
\end{align}
Though simpler than the full MRSSM potential, this toy potential has 
all the important features; a Higgs boson $h$ and a scalar $a$ that couples 
to a ``$D$-term'' $g\,h^2$.   We can study this potential in three
interesting limits:  If $\lambda\mu = m_S = 0$, the field $a$ 
can be integrated out exactly and the residual potential is purely 
quadratic in $h$.  
If $\lambda\mu \ne 0,\ m_S = 0$, an effective Higgs quartic is generated
\begin{eqnarray}
\lambda_{eff} &=& \frac{g\,\lambda\mu}{M_D} 
                  - \frac{\lambda^2\,\mu^2}{4\,M_D^2} 
                  + \mathcal{O} \left( \frac{1}{M_D^3} \right) \, . 
\end{eqnarray}
Notice that the quartic receives a positive contribution 
(if $\lambda \mu > 0$) to order $\mathcal{O}(1/M_D)$, 
and a negative contribution to order $\mathcal{O}(1/M_D^2)$.
We will see the same result in the MRSSM\@, suggesting a modest
hierarchy with $\mu \ll M_D$ maximizes the quartic coupling.
If $m_S \ne 0, \lambda\mu = 0$ (while assuming $m_S \ll M_D$), 
we get a different effective Higgs quartic
\begin{eqnarray}
\lambda_{eff} &=& g^2 \left( \frac{m^2_S}{M_D^2} \right) 
                  + \mathcal{O} \left( \frac{1}{M_D^4} \right) \, .
\end{eqnarray}
Finally, if both $\lambda\mu \ne 0, m_S \ne 0$, the effective quartic is the 
sum of the last two equations. We can remove the quadratic term by 
demanding a minima at $h = v$; 
the resulting potential is then entirely proportional to $\lambda_{eff}$, 
and therefore  $\propto 1/M_D^n, n \ge 1$. Explicitly,
\begin{eqnarray}
V &=& m_0^2\,h^2 + \lambda_{eff} h^4 
      \longrightarrow \lambda_{eff} (h^2 - v^2)^2 \nonumber,
\end{eqnarray}
where we have added an unimportant overall constant.
 
Because the tree-level potential is suppressed, the nature of the 
zero temperature electroweak symmetry-breaking (EWSB) minima of the 
full 1-loop potential is somewhat more complicated. The EWSB minima can be 
nearly degenerate, or even higher than the zero temperature, trivial 
vacua.\footnote{The EWSB conditions used to fix the soft masses 
$m^2_{Hu}, m^2_{Hd}$ only ensure the electroweak vacua is a local minima, 
and not necessarily the global minima.} When the trivial vacuum is 
only slightly higher than the EWSB vacua (at $T=0$), 
the critical temperature will be low, making large $\phi_c/T_c$ easier 
to achieve.  
A cartoon depicting the nearly degenerate minima scenario 
with more typical scenarios is shown below in Fig.~\ref{fig:theveff}.
This effect is mitigated somewhat by the presence of
an effective Higgs potential below the scale of the 
Dirac gaugino masses (needed to obtain a phenomenologically viable Higgs mass).

\begin{figure}[t!]
\includegraphics[width=0.48\textwidth]{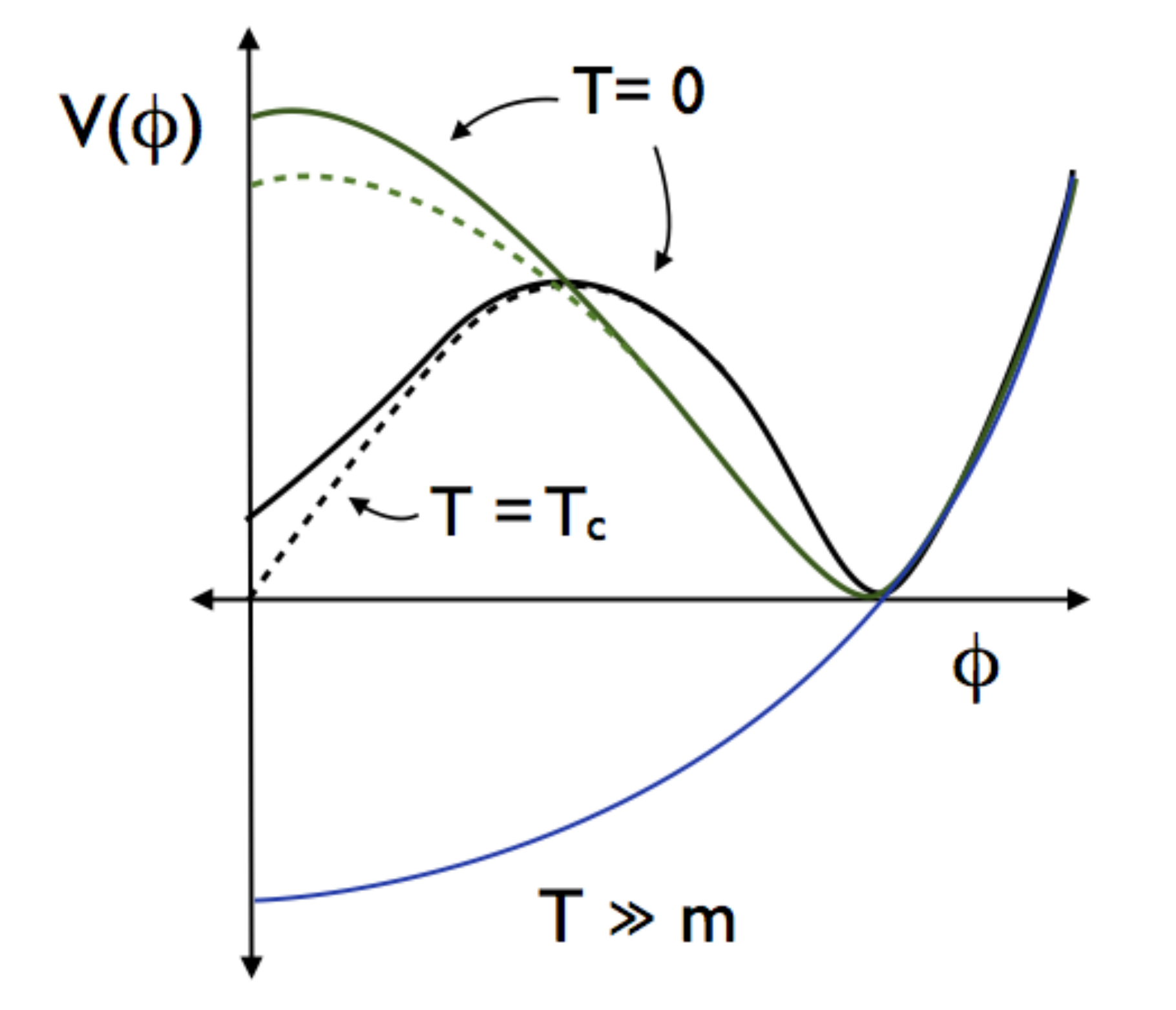}
\caption{Cartoon showing the $T=0$ Higgs potential in a 
nearly-degenerate minima scenario, as in the MRSSM (black solid) 
and a more conventional Higgs potential (solid green). 
The zero of these $T=0$ has been chosen such that $V(\phi_c) = 0$, 
where, roughly, $\phi_c$ is the Higgs vacuum expectation value (vev) at the critical tempterature. 
When the temperature is raised slightly, the potential shifts 
to the dashed lines. For the degenerate case, the small shift in 
temperature brings us to $T_c$, while for the typical potential 
the temperature must be raised higher for a phase transition. 
A high temperature, $T \gg$ all masses, potential is also shown (blue).} 
\label{fig:theveff}
\end{figure}

\section{Effective Potential in the MRSSM}
\label{sec:effpot}

In this section we will describe the various scalar field contributions
from the MRSSM that enter the effective potential at both zero temperature 
and finite temperature.

\subsection{Zero temperature Potential}
\label{sec:tzero}

We will make the following simplifications to avoid an overabundance 
of parameters: (Dirac) gaugino masses $M_1 = M_2 \equiv M$, 
``$\mu$ -terms'' $\mu_u = \mu_d \equiv \mu$, additional Yukawa couplings 
$\lambda^B_u = \lambda^B_d  \equiv \lambda_B$,  
$ \lambda^W_u = \lambda^W_d \equiv \lambda_W$ and 
equal soft-masses for $R_{u,d}$, $m_{R_u} = m_{R_d} \equiv m_R$. 
To satisfy precision electroweak constraints, we take $M$ to be 
much larger than all other scales. 
The neutral scalar components of $\Phi_B$ and $\Phi_W$ are expanded as
\begin{eqnarray}
A_B &=&     \frac{s_0 + i\,p_0}{\sqrt{2}}, \quad A^0_W 
    \; = \; \frac{s_3 + i\, p_3}{\sqrt{2}}
\label{eq:phiscalars}
\end{eqnarray}

The first ingredient in the calculation is the tree-level potential 
$V_{tree}$, which consists of the superpotential piece, the soft masses 
and the $D$-term potential. As we discussed earlier, the interplay 
between the $D$-term and the other contributions results in a 
viable Higgs potential with an EWSB vacuum.

Faced with the hierarchy among superparticle masses ($M \gg \mu$, etc.), 
we proceed by integrating out all particles with mass $M$. 
These include the gauginos (both charged and neutral) and several scalars. 
The scalars with mass $M$ include two CP-even neutral states and 
one charged scalar.  The origin of their mass 
can be traced back to Eq.~(\ref{eq:dmass}): they are, up to 
small mixing effects, the real scalars within $A_B$ and $A^i_W$. 
Removing the heaviest fields, the residual potential now has 
$\mathcal{O}(1/M)$ and $\mathcal{O}(1/M^2)$ suppressed interactions.

The full potential, and the resulting tree-level mass matrices, 
is shown in Appendix~\ref{sec:appa}. The mass matrices in the low-energy 
effective theory are kept field-dependent, meaning we retain all 
$h_u \equiv \phi_u, h_d \equiv \phi_d$ dependence. 
 
Focusing on the CP-even, neutral scalar sector, we next calculate the 1-loop 
Coleman-Weinberg (CW) potential. Working in the $\overline{MS}$ scheme, 
\begin{eqnarray}
V_{CW} &=& \sum_i \frac{n_i}{64 \pi^2} m_i(\phi_u, \phi_d)^4 
           \Bigg( \ln \frac{m_i (\phi_u,\phi_d)^2}{\Lambda^2} - c_i \Bigg) 
           + \delta V, \nonumber \\
\delta V &=& 
- \phi^2_u \left( \frac{\partial V_{CW}}{\partial \phi^2_u} \Big|_v \right) 
- \phi^2_d \left( \frac{\partial V_{CW}}{\partial \phi^2_d} \Big|_v \right) 
\label{eq:cwpo}
\end{eqnarray}
where $\delta V $ is a counterterm that is added to $V_{CW}$ to ensure 
that the 1-loop potential has an extrema at 
$\phi_u = v_u$, $\phi_d = v_d$.  We have made the distinction 
between $\phi_{u,d}$ and $v_{u,d}$ since, at finite temperature, 
the fields will deviate their $T=0$ vacuum values 
(it is also convenient to parameterize the Higgs field dependence 
in terms of $\phi^2 = (\phi^2_u + \phi^2_d)$ and 
$\chi = \text{arctan}(\phi_u/\phi_d)$). 
The sum in Eq.~(\ref{eq:cwpo}) runs over the relevant (light) particles, 
with $n_i$ counting the degrees of freedom, $m\,(\phi_u, \phi_d)$ 
representing the field-dependent mass, and $c_i$ is a constant equal 
to $3/2$ for fermions and scalars and $5/6$ for gauge bosons. 
All calculations have been performed in Landau gauge
(for recent discussion on gauge-dependent artifacts, 
see Ref.~\cite{Patel:2011th}).  

We will come to the exact states included in the sum shortly, 
however note that the field content is slightly different than in the MSSM. 
The sum in the MRSSM contains no gauginos, but includes all 
$R$-partner fields (both the scalar and fermionic components). 
To simplify the calculation, we will neglect sleptons, first and 
second generation squarks, and the sbottoms since their couplings 
to the Higgs are small. The final parameter in the CW potential 
is the renormalization scale $\Lambda$. In order to minimize the effects from higher order terms, we take $\Lambda$ equal to the mass of the heaviest dynamical field, $\Lambda = \text{max}(m_i(\phi))$~\cite{Morrissey:2012db}.

Before moving to finite temperature, the total (tree + 1-loop) 
$T= 0$ potential must satisfy several consistency checks. 
First, the EWSB minima must be the lowest minima in order for 
the vacuum to be stable rather than meta-stable. 
\begin{eqnarray}
V_{T=0}(\phi_u = v_u, \phi_d = v_d) &<& V_{T=0}(\phi_u = \phi_d = 0)
\quad 
\end{eqnarray}
While this condition is always applied, it has little impact on 
the parameter space of models with an unsuppressed tree-level potential. 
The second condition is that the EWSB is a minimum and not a saddle point. 
The counterterms added to $V_{CW}$ only require the vacuum values 
$\phi_u = v_u, \phi_d = v_d$ extremize the potential. 
To assure a minimum, we must also enforce
\begin{eqnarray}
\mathrm{det} \left. \left( 
           \frac{\partial^2 V_{T=0}}{\partial \phi_i \partial \phi_j} 
           \right) \right|_v\, &>& 0 \qquad i = u,\, d \; .
   \label{eq:hessian}
\end{eqnarray}
This condition is automatically satisfied so long as the
Higgs boson masses are positive.  

The effects of the various fields on the Higgs potential clearly depends 
on their mass and spin. Under the assumption that $M \gg$ 
all other mass scales, the mass eigenvalues fall into several categories:

{\bf Category 1:}  
The first category contains particles of mass $\sim M$; 
very heavy fields that we have already integrated out.

{\bf Category 2:}
The second category contains lighter fields that have $\phi$-independent mass. 
These fields shift the Higgs potential only by an overall constant 
and are therefore unimportant to our calculation of the phase transition. 
The higgsinos fall into this category, as do a full multiplet of 
Higgs scalars (charged, neutral CP-even, neutral CP-odd). 
The Higgs multiplet in this category corresponds roughly the 
$H^0, A^0$ and $H^{\pm}$ of the usual MSSM\@. Because of our assumption 
that the $R_{u,d}$ scalars have a common mass, one of the neutral scalars 
also has a $\phi$-independent mass.

{\bf Category 3:}
The third category contains fields with mass of the form $m_0 + f(\phi)$, 
where $m_0$ is a weak scale parameter 
($\mu$, or one of the soft masses other than $M$) while the 
Higgs field-dependence is an additive function $f(\phi)$. 
The remaining charged Higgs scalars, 
the imaginary parts of the $A_B, A^i_W$ scalars, the charged $R$-Higgs scalars, 
and one of the neutral $R$-Higgs scalars all have masses of this type. 
Being light and with $\phi$-dependent masses, these states are 
especially relevant for us, so their masses are explicitly displayed 
below using $\phi_u = \phi\,\sin{\chi},\, \phi_d = \phi\,\cos{\chi}$:
\begin{eqnarray}
m^2_{H^{\pm}_3} &=& 
    m^2_{st} - m^2_{pt} + \frac{\lambda_W^2}{8}\,\phi^2 
    \\
m^2_{A^0_3} &=& \Delta_A 
    + \frac{\lambda_B^2 + \frac 1 4 \lambda_W^2}{4}\,\phi^2 
    \nonumber \\
    & &{} - \sqrt{(\lambda_B^2 + \frac{1}{4}\lambda_W^2)^2\,\phi^4 
          + \Delta'_A(\lambda_B^2 - \frac{1}{4}\lambda_W^2)\Delta'^2_A \phi^2} 
    \nonumber \\
m^2_{A^0_4} &=& \Delta_A 
    + \frac{\lambda^2}{2}\,\phi^2 
    \nonumber \\
    & &{} + \sqrt{(\lambda_B^2 + \frac{1}{4}\lambda_W^2)^2\,\phi^4 
          + \Delta'_A(\lambda_B^2 - \frac{1}{4}\lambda_W^2)\Delta'^2_A \phi^2} 
    \nonumber \\
m^2_{Ru^0} &=& 
    \mu^2 + m^2_R + \lambda^2\,\phi^2 \nonumber \\
m^2_{Ru^-} &=& 
    \mu^2 + m^2_R + \lambda^2\,\phi^2\,\sin^2{\chi} \nonumber \\
m^2_{Rd^+} &=& 
    \mu^2 + m^2_R + \lambda^2\,\phi^2\,\cos^2{\chi}  \nonumber \\
\Delta_A &\equiv& m^2_{s0} + m^2_{st} - m^2_{p0} - m^2_{pt} \nonumber \\
\Delta'_A &\equiv& m^2_{s0} - m^2_{st} - m^2_{p0} + m^2_{pt} \nonumber,
\label{eq:lightmasses}
\end{eqnarray}
where the soft masses $m^2_{s0}, m^2_{p0}, m^2_{st}, m^2_{pt}$ 
for the scalars in Eq.~(\ref{eq:phiscalars}) are 
defined in the Appendix in Eq.~(\ref{eq:softmm}). 

{\bf Category 4:}
The fourth and final category contains fields whose mass comes 
entirely from electroweak symmetry breaking.
This includes the weak gauge bosons and the top quark.
If certain combinations of the $A_B, A^i_W$ soft masses (i.e. $\Delta_A$) happen to be small,
one or more of $H^{\pm}_3,\, A^0_3,\, A^0_4$ will also receive their mass entirely from electroweak symmetry
breaking.

\subsection{Lightest Higgs Mass}
\label{sec:lightmh}

The lightest Higgs boson mass in the MRSSM deserves special attention.
In general, it receives three main contributions to its mass 
to one-loop:
\begin{eqnarray}
m_h^2 &=& m_{h,{\rm tree}}^2 
          + \delta m_{h,\tilde{t}}^2 
          + \delta m_{h,\lambda}^2 \; .
\end{eqnarray}
Unlike the MSSM, there are no tree-level contributions from
the usual $D$-term \cite{Fox:2002bu}.  This would-be disaster
is averted in the MRSSM due to new tree-level contributions from 
the $\lambda$-terms as well as soft-mass contributions to 
the adjoint scalars.  
The general expression can be straightforwardly evaluated
numerically from the effective potential, which we do in
our numerical results below.
The leading contributions, to $\mathcal{O}(1/M^3)$,
can be obtained analytically in the limits 
$|\mu/M|, \, v/|M|, \, m_S^2/M^2 \ll 1$, and $\tan\beta \gg 1$:
\begin{eqnarray}
m_{h,{\rm tree}}^2 &=& 
  \frac{v}{\sqrt 2 M}  M_Z \mu (\lambda_W \cos\theta_W - 2 \lambda_B \sin\theta_W)
  \nonumber \\
& + & \frac{v^2}{32 M^2} \Big[ 
  4 \lambda_B^2 (3 M_Z^2 \sin^2\theta_W - 4 \mu^2) \nonumber \\
& &{} \qquad\;\; - \lambda_W^2 (4 \mu^2 + 3 M_Z^2 \cos^2\theta_W) 
      \nonumber \\
& &{} \qquad\;\; - 6 \lambda_B \lambda_W M_Z^2 \sin^2\theta_W \Big] 
      \nonumber \\
& + & M_Z^2 \frac{\sin^2\theta_W (m_{s0}^2 + m_{p0}^2) 
                    + \cos^2\theta_W (m_{st}^2 + m_{pt}^2)}{4\,M^2} 
      \nonumber \\
& + & \mathcal{O}\left( \frac{1}{M^3} \right)
\label{eq:mhlite}
\end{eqnarray}
The tree-level contribution is maximized when $\lambda_W \mu > 0$
simultaneously with $\lambda_B \mu < 0$, with a phase convention 
where the Dirac gaugino masses are real and positive. 

Taking $\lambda \equiv \lambda_W = -\lambda_B$,
$m_{s0} = m_{p0} = m_{st} = m_{pt} \equiv m_S$, and evaluating 
the contributions for characteristic masses that we will see
later in our numerical evaluation:
\begin{eqnarray}
m_{h,\mathrm{tree}}^2 & \simeq & 
      0.7 M_Z^2 \lambda \frac{\mu}{200 \; \mathrm{GeV}} 
                \frac{1 \; \mathrm{TeV}}{M} \nonumber \\
& &{} - 0.18 M_Z^2 \lambda^2 
      \left( \frac{\mu}{200 \; \mathrm{GeV}} \right)^2
      \left( \frac{1 \; \mathrm{TeV}}{M} \right)^2 \nonumber \\
& &{} + 0.02 M_Z^2 \lambda^2 \left( \frac{1 \; \mathrm{TeV}}{M} \right)^2 
      \nonumber \\ 
& &{} + \frac{1}{2} M_Z^2 \frac{m_S^2}{M^2} \; .
\end{eqnarray}
This approximate expression slightly underestimates the tree-level 
contribution.  Nevertheless, we see that we can achieve 
$m_{h,\mathrm{tree}}^2$ nearly equal to $M_Z^2$ with $\lambda \simeq 2$,
intriguing (though accidentially) similar to the largest tree-level 
value found in the MSSM\@.

The one-loop contributions to the lightest Higgs mass from the stops 
$\delta m_{h,\tilde{t}}^2$ are well-known~\cite{Martin:1997ns} and won't be 
repeated here.  We note, however, that there is no scalar
trilinear coupling $A_t \tilde{t}_L \tilde{t}_R^* h$ due to $R$-symmetry, 
and thus $A_t = 0$. This means we will need stops
with masses above a few TeV to obtain a large enough one-loop 
radiative correction to the Higgs mass, and as a consequence, 
we can integrate the stops out in the calculation of the electroweak 
phase transition.

Finally, there are additional 
one-loop contributions to the Higgs mass 
from the terms proportional to $\lambda$.
These are straightforwardly derived from the Coleman-Weinberg
potential.

\subsection{Finite Temperature Contributions}
\label{sec:gettc}

The effective potential at $T>0$ can be separated into a 
temperature-independent contribution as well as a temperature-dependent 
contriubtion. 
The temperature-independent contribution is the tree-level plus 
Coleman-Weinberg potential 
calculated in the previous section.  The temperature-dependent
contribution includes 
\begin{eqnarray}
V_{T} &=& n_i \frac{T^4}{2\pi^2}\, J_i(m^2_i(\phi_u, \phi_d)/T^2), \\
      J_{\pm}(y) &=& \int_0^{\infty} dx\, x^2\, \nonumber
          \log{\left[1 \pm \exp\left(-\sqrt{x^2 + y^2} \right) \right]},
\end{eqnarray}
where $J_+ (J_-)$ is the thermal function for fermions (bosons), 
respectively. The thermal potential must be amended due to some 
well-known subtleties of perturbation theory in finite temperature, 
however before addressing these it is important to break down the 
effects of moving to $T \ne 0$.

In the limits $T \gg m$ and $T \ll m$ the thermal functions have 
a simple form
\begin{eqnarray}
V_T = \left\{ \begin{array}{cc} \frac{-|n_i|T^4\,\pi^2}{90} & T \gg m \\ -|n_i|\,T^4\, \Big( \frac{m^2}{2\pi\,T^2} \Big)^{3/2}\,e^{-m/T} & T \ll m \\ \end{array} \right.
\label{eq:thermal}
\end{eqnarray}

In light of these limits, finite temperature effects from particles 
with mass $M \gg T$ are completely negligible. Similarly, fields 
with purely $\phi$-dependent mass have the largest impact on the 
shape of the potential. For $\phi = 0$, these fields are light 
(up to thermal masses, which we will come to shortly), 
so the thermal contribution is negative definite and $\propto T^4$, 
meanwhile, out at larger field values (of $\phi$), all these fields 
are heavy so the thermal corrections are small. 
Thus, fields whose mass comes entirely from $\phi$ push the trivial 
($\phi = 0$) vacuum downwards sharply as the temperature increases 
while leaving the large-$\phi$ part of the potential unaffected. 
Because the thermal corrections at $\phi = 0$ depend so strongly on $T$, 
the more degrees of freedom in the $m \propto \phi$ category, 
the lower we need to raise the temperature before the trivial vev 
and EWSB vev equilibrate, leading to larger $\phi_c/T_c$. 
For fields with mass of the form $m = m_0 + f(\phi)$ the 
thermal contribution depends in detail on the relation between 
$m_0$ and $T$, so we must evaluate these contributions
numerically. 

\subsubsection{Thermal masses}

Thermal masses are systematically calculated by summing over the 
``daisy'' diagrams~\cite{Parwani:1991gq,Carrington:1991hz,Arnold:1992rz} 
where the contribution to their mass is typically of order $g^2\,T^2$. 
Physically, they represent the screening of 
scalar fields in a thermal bath. Their effect is to reduce the 
$\phi$-dependence in scalar field masses, hence weakening the 
phase transition. 

Thermal masses are most important for fields whose mass is determined
entirely from electroweak symmetry breaking ($m \propto \phi$),
since these fields become massless as $\phi \ra 0$. 
For fields whose mass has a $\phi$-independent piece, 
the thermal contribution has little effect. 
Therefore, we include thermal masses only for the charged Higgs, 
$H^{\pm}_3$ and the relevant CP-odd charge-neutral Higgs fields $A^0_{3,4}$. 
Thermal masses for the longitudinal $W/Z$ degrees of freedom 
($\mathcal{O}(g^2)$) and all other scalars (in categories 1-3) 
are neglected for simplicity.

We evaluate the thermal mass correction in the interaction basis. 
We use the expressions in Ref.~\cite{Comelli:1996vm}, 
ignoring terms proportional to the electroweak gauge couplings 
and only retaining terms of order $\lambda^2\, T^2$, 
with $\lambda \sim \mathcal{O}(1)$. In the large gaugino mass limit, 
the only fields we need to consider are the $\im(A_B)$ and the $\im(A^i_W)$; 
the light pseudoscalars are combinations of $\im(A_B)$ and $\im(A^3_W)$, 
and the charged Higgs fields are made up of $\im(A^{\pm}_W)$. 
For this subset of fields, the thermal mass enters as additional terms 
in the interaction basis mass matrix $\Pi_{ij}$, where $i,j$ runs over 
$(\im(A_0), \im(A_3))$ ($(\im(A^+),\im( A^{-*})) + c.c.$) for the 
CP-odd Higgs 
(charged Higgs). Only self energy diagrams giving rise to a 
quadratic divergence at $T=0$ contribute to the thermal mass 
(i.e.\ requires $1/p^2$ in the integrand of the loop integral 
by power counting).  Therefore, only self energy diagrams originating 
from the 4-point scalar interactions and scalar-fermion-fermion 
interactions without interior chirality flips contribute. 
By this argument, supersoft interactions do not contribute to the 
thermal masses as they contain only 3-point scalar interactions. 
Similarly, $D$-term contributions are proportional to the 
electroweak gauge couplings and they are negligible in comparison 
to $\lambda$.  The only interaction terms that can contribute come 
from Eq.~(\ref{eq:extrayuk}). However, the interactions in 
Eq.~(\ref{eq:extrayuk}) do not generate off-diagonal 
thermal masses -- the correction to the $\im(A_0)-\im(A_3)$ entry 
is proportional to $\sum T_3 = 0$, and there no sufficiently 
strong/divergent interactions to create a $\im(A^+) \, \im(A^{-,*})$ entry. 
The diagonal elements 
$\Pi_{A^0 A^0}, \Pi_{A^3 A^3}, \Pi_{A^+ A^-}, \Pi_{A^{+*} A^{-*}}$, 
on the other hand, are non-zero:
\begin{eqnarray}
\Pi_{A^0 A^0}  &=& \frac{1}{2} \lambda_B^2 T^2 \\ 
\Pi_{A^3 A^3}  &=& \frac{1}{8} \lambda_W^2 T^2 \\ 
\Pi_{A^+ A^-}  &=& \Pi_{A^{+*} A^{-*}} \; = \; \frac{\lambda_W^2}{4}  T^2 \, .
\end{eqnarray}

In the mass basis, 
\begin{eqnarray}
\bar{m}_{a^0}^2     &=& m_{a^0}^2 + \frac{1}{4} 
                        (\lambda_B^2 + \frac{1}{4}\lambda_W^2) T^2 \\
\bar{m}_{h^{\pm}}^2 &=& m_{h^{\pm}}^2 + \frac{\lambda_W^2}{4} T^2
\end{eqnarray}

To incorporate the thermal masses, we follow the procedure in 
Ref~\cite{Parwani:1991gq,Carrington:1991hz,Arnold:1992rz} 
and include a ``ring'' contribution
\begin{eqnarray}
V_{ring} &=& \frac{T}{12\pi} \sum_i^{scal.} \, n_i \, 
    (m^3_i(\phi_u, \phi_d) - \tilde m^3_i(\phi_u, \phi_d) ), \nonumber \\
         & & \tilde{m}^2_i(\phi_u, \phi_d) 
             \, = \, m^2_i(\phi_u, \phi_d) + \Pi_i,
\end{eqnarray}
where $i$ runs over $\{A^0_3, A^0_4, H^{\pm}_3\}$.

\section{Phase Transition: Numerical Results}
\label{sec:numresults}

Over the last few sections, we have discussed the components of the complete, 
finite temperature plus 1-loop potential
\begin{eqnarray}
V &=& V_{tree} + V_{CW} + V_T + V_{ring}.
\end{eqnarray}

We are now ready to assemble the pieces and begin our hunt for regions where the phase transition is strong. Starting with the tree-level Higgs potential, we integrate out all heavy fields; for the mass $\sim M$ scalars and gauginos, this is done at tree level, while the stop squarks must be integrated out at one-loop level. This potential, restricted to light fields, is then augmented by the Coleman-Weinberg potential. We only include states in the Coleman-Weinberg that are light and whose couplings to the Higgs are unsuppressed -- specifically, SM gauge bosons, the top quark, and the six states in Eq.~(\ref{eq:lightmasses}).  The CW contributions from $\phi-$independent states merely shift the effective potential by an additive constant and are uninteresting. For a given parameter set, the scale $\Lambda$ in the CW potential is set to $\text{max}(m_i(\phi))$ to minimize the effect of higher order corrections. The finite temperature potential -- for the same set of states as we used in the CW piece --  is then added, along with ring contributions for any bosonic fields whose mass is directly proportional to $\phi$ (i.e. gauge bosons).

As in Sec.~\ref{sec:effpot}, we will make several assumptions in order to reduce the parameter space for our numerical study:
\begin{align} 
M_1 = &\, M_2 \equiv M \nn \\
\mu_u = &\, \mu_d \equiv \mu \nn \\
\lambda^u_B = \lambda^d_B \equiv \lambda_B ,&\, \lambda^u_W = \lambda^d_W \equiv \lambda_W \\
m_{R_u} = &\,m_{R_d} \equiv m_R \nn  \\
m_{s0}^2 = m_{st}^2  = &\, m_{p0}^2 = m_{pt}^2 \equiv m_S^2 \nn 
\label{eq:numass}
\end{align}
The first five assumptions are identical to those in Sec.~\ref{sec:effpot}. The final assumption, the equality of the $F$-term masses for the $A_i$ scalars sets $\Delta_A = \Delta'_A = 0$ in Eq.~(\ref{eq:lightmasses}), removing all $\phi$-independent contributions to $m^2_{H^{\pm}_3}$ and $m^2_{A^0_4}$ (in terms of the mass categories laid out earlier, these two states move from category three to category four). Under these assumptions, the strength of the phase transition is a function of eight parameters: $\tan{\beta}, m_A, M, \mu, \lambda_B, \lambda_W, m^2_S$ and $m_R$. However, since $m_R$ always appears with $\mu$ in the combination $\mu^2 + m^2_R$, we will set $m_R = 0$ and use $\mu$ as a proxy for the two, reducing the problem to seven parameters. 

Requiring the gaugino mass $M \ge 1$ TeV, we scan over the other parameters, evaluating $\phi_c, T_c$ and the lightest Higgs mass at every point. Starting with the full Higgs potential raised to a sufficiently high temperature ($\sim$250 GeV), we lower the temperature in successively smaller steps until we find a minima in the potential that is degenerate with the electroweak symmetric minima. In principle we ought to search for minima in a two-dimensional space: ($\phi_u, \phi_d$) or, equivalently $(\phi, \tan{\chi})$. However, in performing this two-dimensional search, we find that the ratio of the Higgs fields at $T_c$ remains close to the vacuum value (the ratio of vevs) provided $\tan{\beta}$ is $\gtrsim 3$. Therefore, to more quickly scan over several parameters, we focus on $\tan{\beta} \gtrsim 3$, set $\tan{\chi} \equiv \tan{\beta(T_c)} = \tan{\beta} \equiv \tan{\beta(T = 0)}$, effectively reducing the Higgs potential to a function of $\phi$ alone.

With $M$ and $\tan{\beta}$ essentially fixed by our approximations, our parameter set is reduced to $m_A, \mu, \lambda_B, \lambda_W$ and $m^2_S$. Of these, $m_A$ and $m^2_S$ have little impact on the strength of the phase transition; both appear only in the tree-level Higgs potential (given our assumptions in Eq.~(\ref{eq:numass})), either as $\phi$-independent terms or suppressed by two powers of the heavy scale $M$. Therefore, we show our results by fixing $m_A = 300\, \gev, m_S = 0$ and scanning in various directions over the most sensitive parameters, $\mu, \lambda_B, \lambda_W$.

\begin{figure*}[t!]
\includegraphics[width=0.48\textwidth]{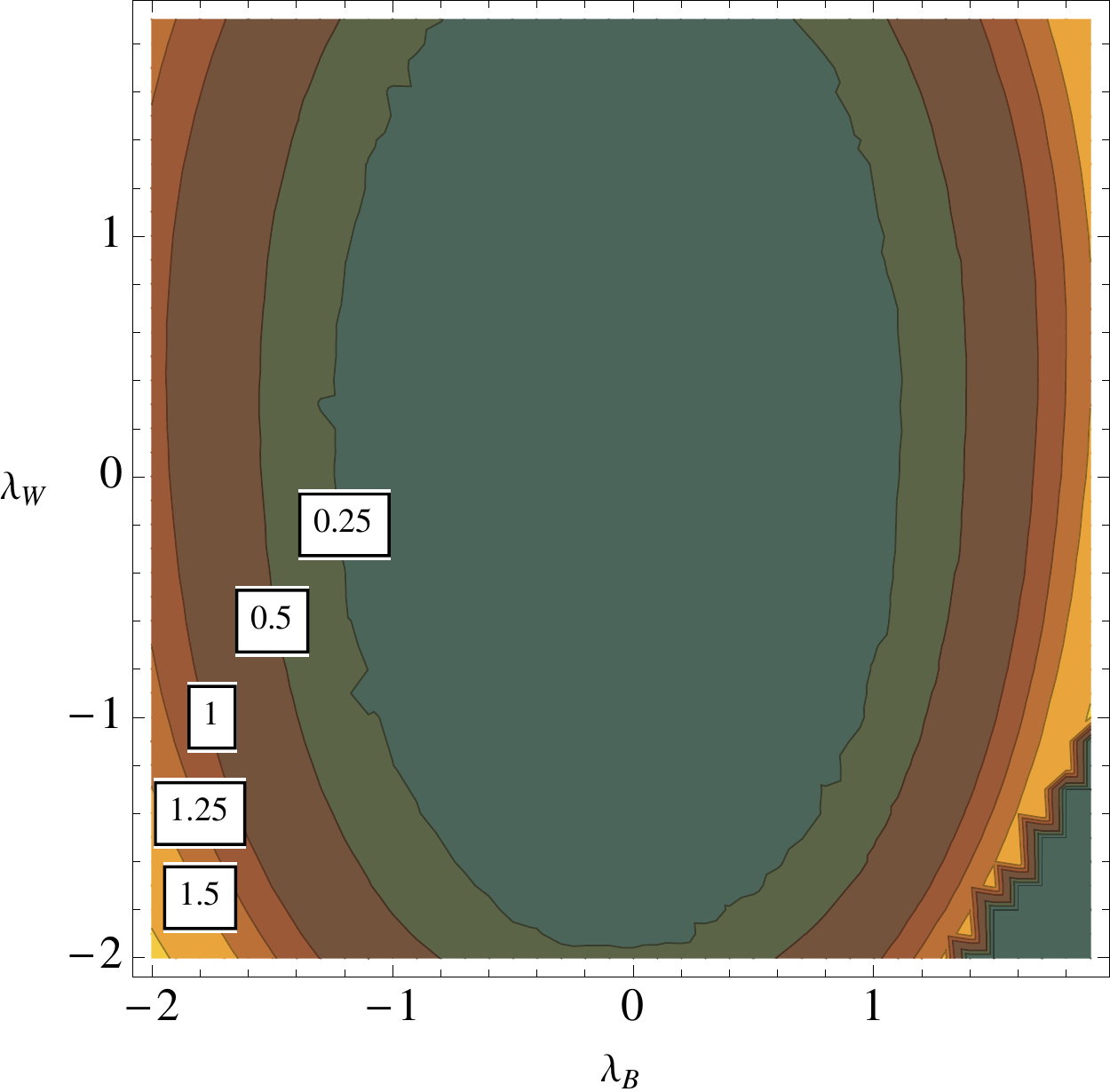}
\hfill
\includegraphics[width=0.48\textwidth]{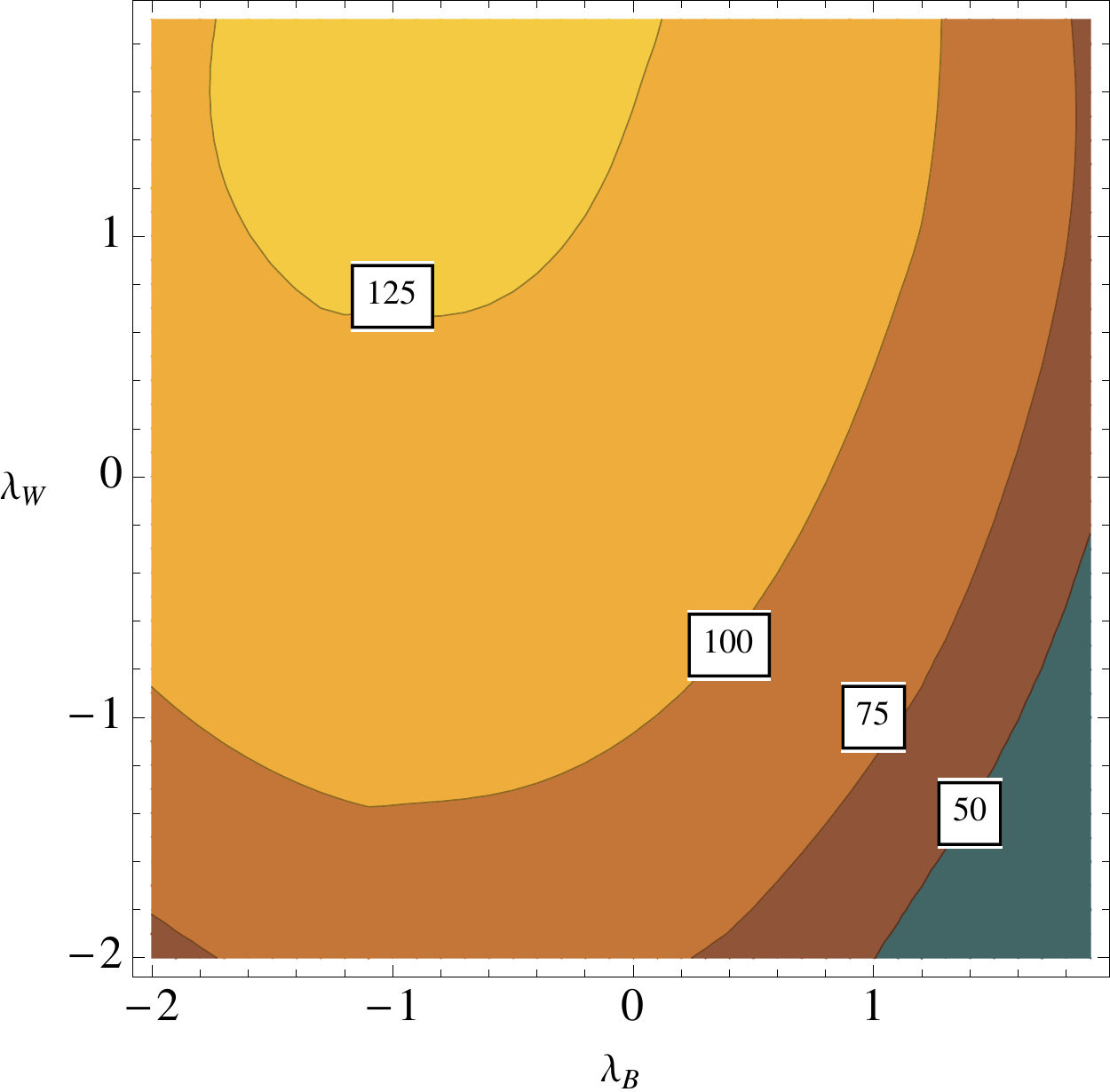}
\caption{On the left, we show contours of 
$\phi_c/T_c$ in the $(\lambda_B,\lambda_W)$ plane 
with $\mu = 200$~GeV and $M = 1$~TeV\@.  The other parameters chosen 
were $\tan\beta = 4$, $m_A = 300$~GeV, $m_s = 0$~GeV, 
but the values of $\phi_c / T_c$ are not particularly sensitive 
to these choices.  On the right, we show contours of the lightest
Higgs mass (in GeV) assuming $m_{\tilde{t}_L} = m_{\tilde{t}_R} = 3$~TeV\@. 
Large stop masses were necessary to obtain contours that approach
$\mh = 125$~GeV\@. \\
\\
}
\label{fig:bullseye}
\end{figure*}

Our first scan, shown in Fig.~\ref{fig:bullseye}, shows the dependence on the $\lambda_i$ couplings for fixed $\mu$. This particular scan was performed using $M = 1\, \tev$, $\tan{\beta} = 4$, though increasing to larger values for either parameter has a negligible impact. Within the $(\lambda_B, \lambda_W)$ space, the $\phi_c/T_c$ contours trace out a ``bullseye'' shape. This is expected -- the largest effect on the strength of the phase transition should come when the interactions between the Higgs fields and new scalars are strongest. The difference in normalization between $\lambda_B$ and $\lambda_W$, along with factors of $\tan{\theta}$, set how fast the contours change in $\lambda_B$ versus $\lambda_W$. 

The second panel of Fig.~\ref{fig:bullseye} shows the lightest Higgs mass in the same parameter space. Clearly, the region of $\lambda_B < 0, \lambda_W > 0$ is favored since it allows $m_h \gtrsim 125\,\gev$. The shape of the Higgs contours is driven by the $\lambda_i$ dependence of the  tree-level piece, which we explored in Sec.~\ref{sec:tzero}. Loop level contributions to $m^2_h$, though sizable, do not prefer a given sign for the $\lambda_i$ since they are always proportional to $\lambda^2_B, \lambda^2_W$. We emphasize that the relative sign of $\lambda_B, \lambda_W$ is, of course, convention dependent. Specifically, the signs of superpotential couplings depend on the ordering of the fields in Eq.~(\ref{eq:newmu}, \ref{eq:extrayuk}).

\begin{figure*}[t!]
\includegraphics[width=0.48\textwidth]{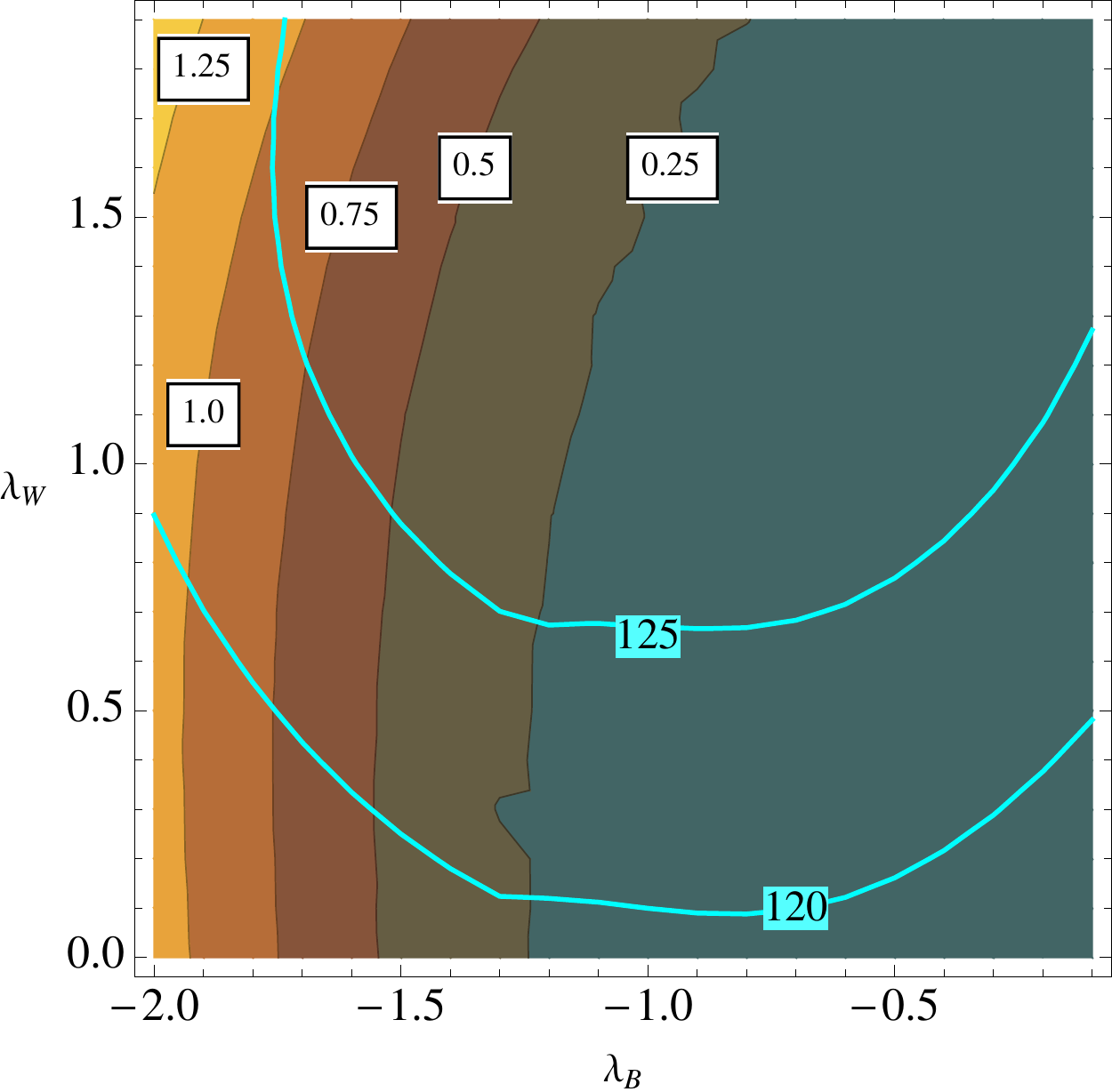}
\hfill
\includegraphics[width=0.48\textwidth]{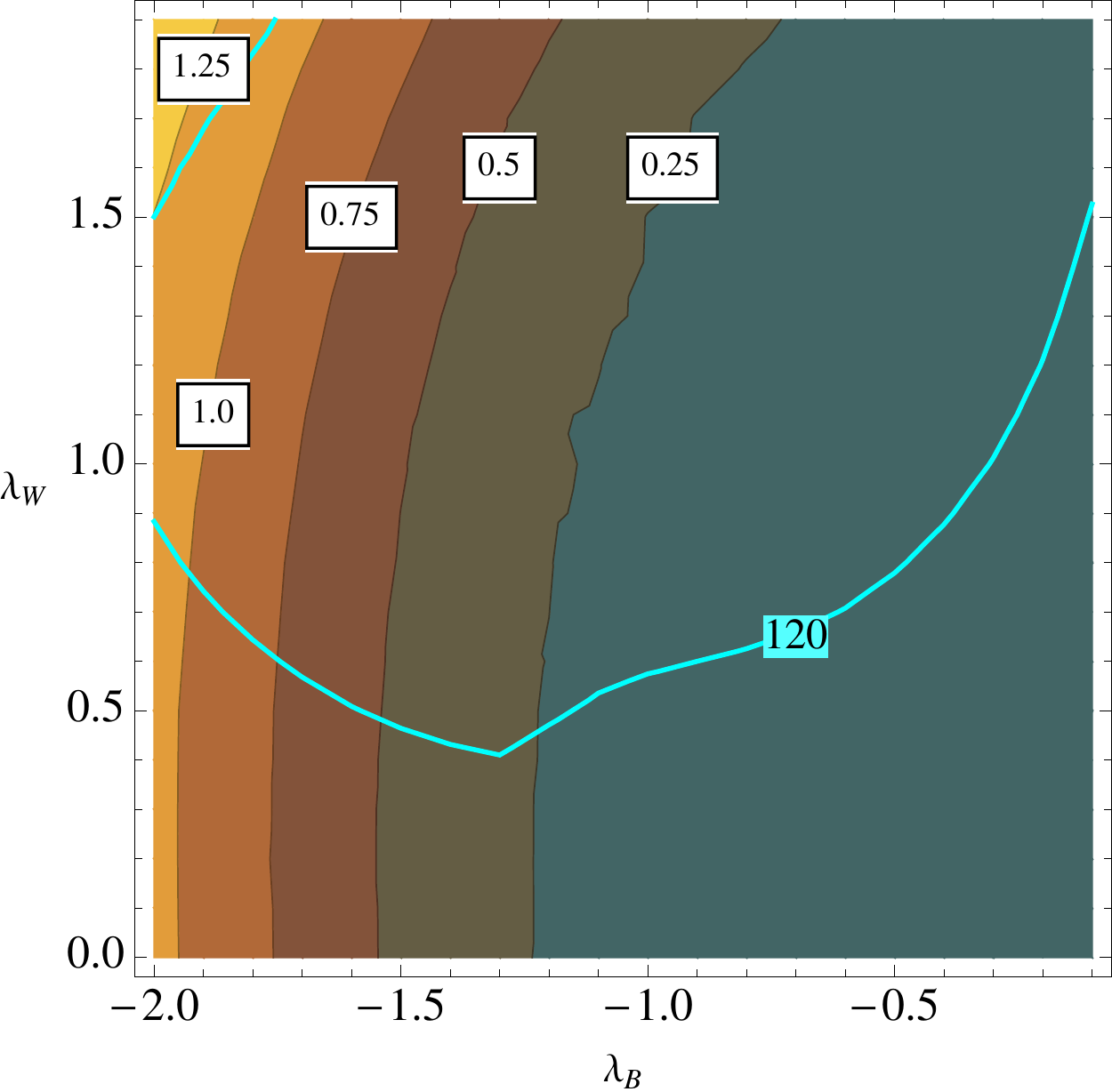}
\caption{Same as upper-left part of Fig.~\ref{fig:bullseye}, 
except we have ``zoomed in'' to the contour regions of interest.
The left-side plot has $M = 1$~TeV, as in Fig.~\ref{fig:bullseye},
while the right-side plot has $M = 2$~TeV; all of the other
parameters are the same as given in Fig.~\ref{fig:bullseye}.
Here we have overlaid Higgs mass (in GeV) contours to demonstrate the
region of overlap for $\mh \simeq 125$~GeV simultaneous
with $\phi_c/T_c \gsim 1$.}
\label{fig:zooms}
\end{figure*}

In Fig.~\ref{fig:zooms}, we zoom in on the most interesting quadrant, $\lambda_B < 0, \lambda_W > 0$, overlaying the lightest Higgs mass contours on top of the $\phi_c/T_c$ contours. To demonstrate the effect of the Dirac gaugino mass, we repeat this zoomed-in scan for a second Dirac mass, $M = 2$~TeV\@. For larger $M$, the strength of the phase transition is hardly changed, while the mass of the lightest Higgs is slightly reduced since the tree-level contribution to $m^2_h$ scales as $1/M$. In both panels of Fig.~\ref{fig:zooms} we can see that there are regions where the Higgs mass is close to $125$~GeV and the phase transition is strong, $\phi_c/T_c \gsim 1$. 

\begin{figure*}
\includegraphics[width=0.32\textwidth]{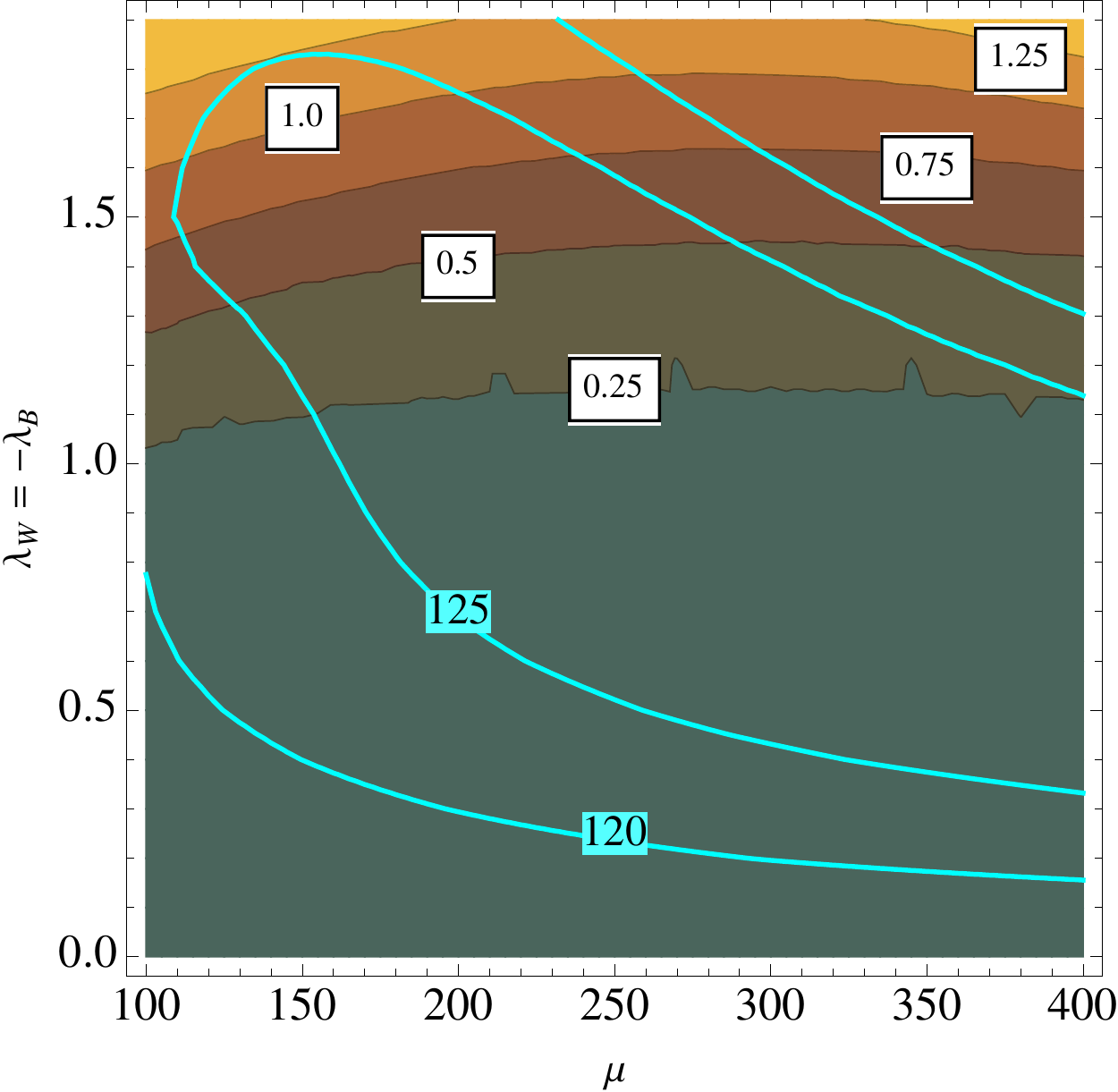}
\includegraphics[width=0.32\textwidth]{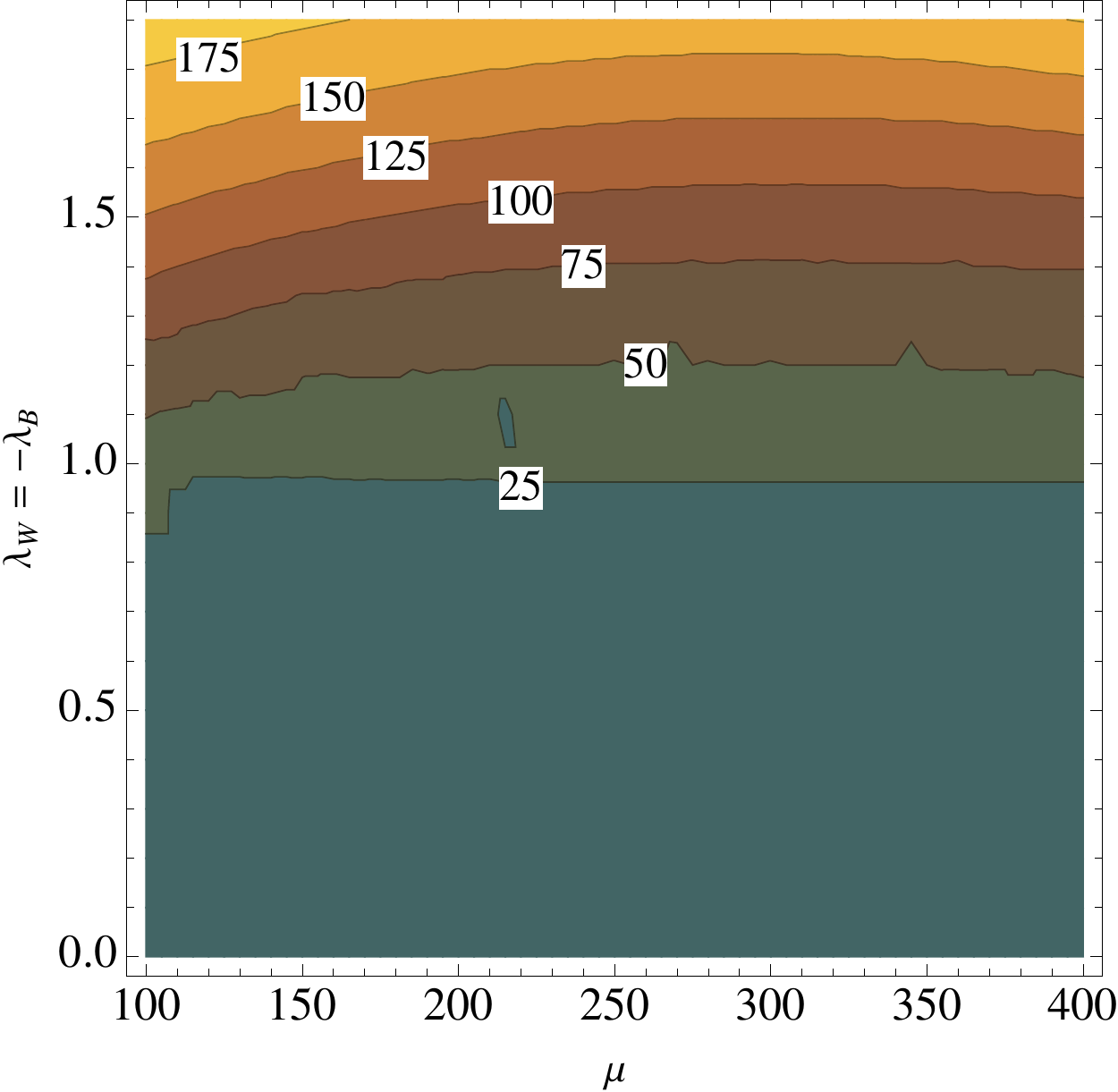}
\includegraphics[width=0.32\textwidth]{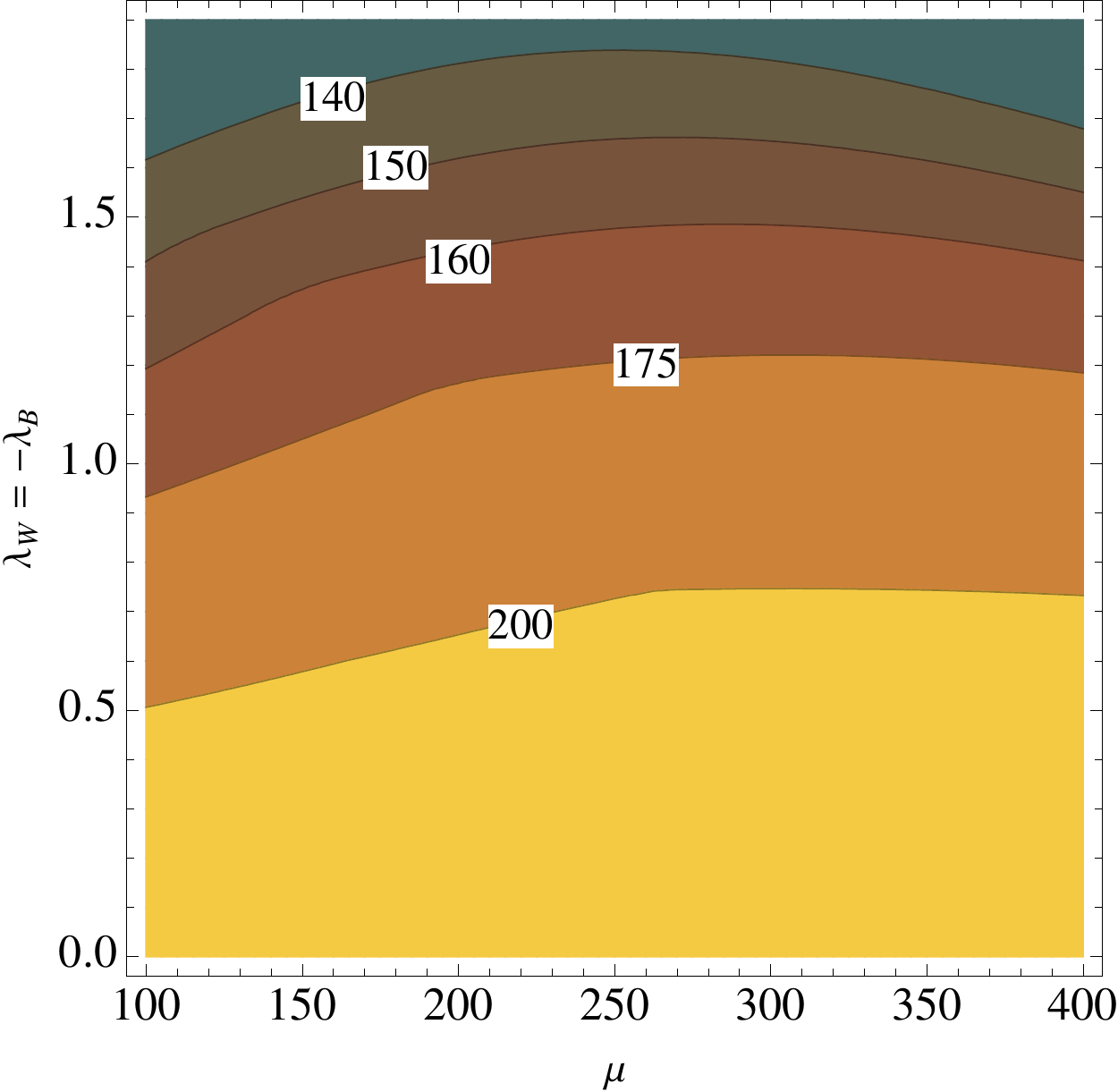}
\caption{Contours of $\phi_c / T_c$ (left panel)
$\phi_c$ (in GeV, middle panel), $T_c$ (in GeV, right panel) in the $(\lambda,\mu)$ plane.
In these figures we have taken $\lambda \equiv \lambda_W = - \lambda_B$
and $\mu = 200$~GeV, $M = 1$~TeV, $\tan\beta = 4$, $m_A = 300$~GeV, 
and $m_s = 0$~GeV\@.}
\label{fig:lammuplot}
\end{figure*}

To study how the strength of the phase transition depends on $\mu$, we consider a different direction in parameter space, where $\lambda_B = -\lambda_W$ while we scan over $\lambda_W$ and $\mu$. The resulting $\phi_c/T_c$ and $m_h$ contours (overlaid) are shown in Fig.~\ref{fig:lammuplot}. For the parameter ranges we have plotted, the $\mu$ dependence of the phase transition is minor. As in Fig.~\ref{fig:zooms}, we see that there are regions where a strongly first order electroweak phase transition can be achieved simultaneously with a $125\,\gev$ Higgs boson. The viable parameter regions all require $\lambda \sim 1.5-2$. Had we extended the scan in Fig.~\ref{fig:lammuplot} to larger values, we would find regions where the condition in Eq.~(\ref{eq:hessian}) is violated because the lightest Higgs mass (squared) is driven negative.

Summarizing our numerical studies, we have shown that the electroweak phase transition can be strong 
over a wide range of viable MRSSM parameters. However, 
a strongly first order phase transition is only part of the 
baryogenesis mechanism. A first order phase transition ensures 
baryon-number-violating sphaleron process are out of equilibrium. 
This prevents sphalerons from erasing any generated baryon asymmetry, 
but we still need to generate an asymmetry in the first place. 
One well-established way to generate an asymmetry is through 
CP-violating collisions between particles in the plasma and the 
walls of vacuum bubbles. We explore this mechanism in the context 
of the MRSSM in the next section.

\section{CP-violation}
\label{sec:cpv}
Tunneling processes can be understood semiclassically by spacetime-dependent field configurations that connect the real and false vacua. For the EW phase transition, the field that interpolates is the Higgs field, $\phi(x)$. Expanding about the tunneling configuration, fields (top quarks, charginos, etc.) that interact with the Higgs appear to have a spacetime-dependent masses; if the Higgs is complex, these masses will also be complex. Complex masses violate CP, so collisions of particles with this spacetime-dependent Higgs can be shown to lead to an asymmetry between particles and antiparticles on either side of the wall~\cite{Cohen:1990it, Cohen:1990py,Farrar:1993sp, Farrar:1993hn}.  A CP asymmetry generated at the outer edge of the bubble wall is then propagated to the interior of the bubble (where sphaleron effects are unsuppressed) via diffusion and transport mechanics~\cite{Cohen:1994ss,Riotto:1995hh,Riotto:1997vy,Riotto:1998zb,Konstandin:2004gy,Huber:2006wf}. Provided a strong enough source of CP violation, the observed baryon asymmetry can easily be generated. The difficulty lies in introducing a CP violating source that is strong enough, once manipulated into the form of a complex Higgs coupling/complex mass, yet secluded enough to avoid conflict with existing flavor and CP observables. Along with the magnitude of the CP-violating phase, the thickness and speed of the bubble are important parameters in determining the size of the asymmetry. 

The most studied MSSM baryogenesis model uses the relative phase between the $\mu$-term and the gaugino mass ($M_2$) as the CP-violation source~\cite{Carena:1996wj}. However, this phase is stringently bounded by EDM measurements~\cite{Chang:2002ex}, and successful MSSM baryogenesis relies on resonance production when $\mu\sim M_2$. In the MRSSM, we have the opposite problem. There is no significant EDM constraint on the CP phase due to the absence of Majorana mass and left-right squark mass mixing~\cite{Kribs:2007ac,Hisano:2006mv}, but there is also no resonance production enhancement as $M_2\gg\mu$ is required by EW precision measurements. A more careful study is therefore necessary to show that the MRSSM can generate enough asymmetry. 

There are several complex parameters in MRSSM: two higgsino mass terms $\mu_u$ and $\mu_d$; three Dirac gaugino masses $M_i$; three holomorphic scalar masses of the adjoints $m_i^2$; the $B_{\mu}$ term; and four $\lambda$ couplings in Eq.~(\ref{eq:extrayuk}), which give 13 complex phases. The phases of seven superfields $H_{u,d}$, $R_{u,d}$, $\Phi_{\tilde{B},\tilde{W},\tilde{g}}$ can remove six of the phases while keeping an $R$-symmetry one. The seven surviving phases can be parametrized as\footnote{See Appendix~\ref{sec:appa} for the relation between the soft masses $m_{st}, m_{s0}$ etc., and holomorphic/non-holomorphic soft masses.}:
\begin{align}
& m_iM_i^*, i=\tilde{B},\tilde{W},\tilde{g} \nn \\
& \mu_u\,M_j\left(\lambda_u^j\right)^*,\, \mu_d\,M_j\left(\lambda_d^j\right)^*,\, j=\tilde{B},\tilde{W}. 
 \end{align}
 Under our simplifying assumptions of equal $\lambda^u, \lambda^d$ couplings, equal electroweak gaugino masses and equal $\mu_i$ terms, the number of phases is reduced to three: $Arg(m_{\tilde{g}}M^*_{\tilde{g}}), Arg(\mu\, M\,\lambda_B), Arg(\mu\, M\,\lambda_W)$.  The strong CP constraint sets a bound $Arg\left(m_{\tilde{g}}M^*_{\tilde{g}}\right)\ll 10^{-7}$, but there are no significant bounds on the $\mu$-term related phases. 

The phases $Arg(\mu\, M\,\lambda_B), Arg(\mu\, M\,\lambda_W)$ can source CP violation in the MRSSM through the higgsino sector. At leading order in $1/M$, the higgsino masses are simply $\mu$. However, after integrating out the mass $M$ gauginos and scalars, the higgsino masses shift to
\begin{eqnarray}
m_{\chi^{\pm}_1} &=& \mu + \frac{g\,\phi^2(z)\,e^{2iq_{\phi}(z)}}{16M}
  \bigg[ - \sin^2{\beta} \left( 2\lambda_B \tan{\theta} + \lambda_W \right) 
  \nonumber \\
& &{} \quad\quad
    + \cos^2{\beta} \left( 2\lambda_B\tan{\theta} 
                           + (1 + 4\sqrt{2} )\lambda_W \right) \bigg] \\
m_{\chi^{\pm}_2} &=& \mu + \frac{g\,\phi^2(z)\,e^{2iq_{\phi}(z)}}{16M}
  \bigg[ \cos^2{\beta} \left( 2\lambda_B\tan{\theta} + \lambda_W \right) \nn \\
& &{} \quad\quad
      - \sin^2{\beta} \left( 2\lambda_B \tan{\theta} 
      + (1 + 4\sqrt{2}) \lambda_W \right) \bigg] \, , \nn 
\end{eqnarray}
where we have added $\hat z$ dependence to quantities that will vary across the bubble wall\footnote{For simplicity we will assume a planar bubble propagating in the $\hat z$ direction.}, and subsumed the relative phases $\mu M \lambda_W, \mu M \lambda_B$ into a single (spacetime-dependent) phase for $\phi(z)$. The phase in the higgsino masses is suppressed by $1/M$, however, as emphasized above, we can consider much larger phases (and phase changes) $\Delta q \sim \pi$ in the MRSSM as there is no EDM constraint.

To get a better estimate of the asymmetry that can be generated from chargino interactions, we rely on the similarity between the higgsino masses in the MRSSM and the higgsino masses in the nMSSM. Specifically, as shown in Ref.~\cite{Huber:2006wf}, in the limit $M_2 \gg \mu$, the dominant source of CP violation in the nMSSM comes from the changing phase of $\mu$. Within this limit and assuming canonical profiles~\cite{Cohen:1993nk} for the  phase and magnitude of $\mu$ across the bubble, Ref.~\cite{Huber:2006wf} found the generated baryon to photon ratio to be:

\begin{equation}
\eta_{10} \simeq 
    c(T_c)
    \frac{\Delta q}{\pi} 
    \frac{1}{l_w T_c}
    \left(\frac{\mu}{\tau T_c}\right)^{3/2}
    \frac{\Delta\mu}{\tau T_c}\,\exp \left[ -\frac{\mu}{\tau T_c} \right], 
\label{eq:nMSSMasymm}
\end{equation}
where $\eta_{10} \equiv 10^{10}\, \eta$, $c(T_c) \simeq 1.6\, T_c/\mathrm{GeV}$,
$\tau \simeq 0.78$, $\mu$ is the amplitude of the $z$-independent part of the higgsino mass, and $\Delta \mu$ ($\Delta q$) is the change in amplitude (phase) of the coordinate-dependent part of $\mu$ across the bubble. The remaining parameters in Eq.~(\ref{eq:nMSSMasymm}) are the thickness of the wall $l_w$ and the critical temperature $T_c$. 

Applying  Eq.~(\ref{eq:nMSSMasymm}) to the MRSSM, we make the following identifications
\begin{align}
\mu_0 & =\mu \\
 \Delta \mu & = \frac{-g\,v^2\,\cos{2\beta}}{4M}(2\,\lambda_B\tan{\theta}  + (1+ 2\sqrt 2) \lambda_W), \nn \\
 \Delta q & = \Delta q_{\phi}
\label{eq:maptomrssm}
\end{align}
where $\Delta \mu$ is derived by summing over the coordinate dependent parts of both chargino masses. The $\Delta \mu$ also includes an extra factor of two to account for the fact that the $\mu \supset \phi^2(z)$ in the MRSSM, while $\mu \supset \phi(z)$ in the nMSSM \cite{Huber:2006wf}, thus the rate of change across the bubble is twice is large.

As an illustration, consider the point $\lambda_W = -\lambda_B = 2$,
$M = 1$~TeV, $\mu = 200$~GeV, $m_A = 300$~GeV, $\tan\beta = 4$, 
and $m_s = 0$~GeV\@.  
The critcal temperature can be read off from
Fig.~\ref{fig:lammuplot} to be $T_c \simeq 135$~GeV\@.
Using these parameters, including only the chargino contributions
to the baryon asymmetry, with $l_w = 10/T_c$ \cite{Huber:2006wf}
and $\Delta q = \pi$, we obtain $\eta_{10} \simeq 4.0$.
This is larger than needed to match the baryon asymmetry
of the Universe, but it trivial to adjust $\Delta q$ 
or other parameters to bring this in line with $\eta_{10} \simeq 1$.
We have not considered possible contributions from Dirac neutralinos,
nor from squarks and sleptons, which would be interesting to study 
in future work.

\section{Collider Limits}
\label{sec:collider}

As expected from phase transition lore and shown explicitly in Sec.~\ref{sec:gettc}, the phase transition becomes increasingly first order when light scalars are present. Therefore, in order to judge how strong the phase transition can actually be in the MRSSM, we need to know just how light the relevant scalars can be. We begin with a recap of the light particles in this scenario (full mass matrices and eigenstates can be found in Appendix~\ref{sec:appa}):

\begin{itemize}
\item the lightest neutral Higgs boson (other 
      Higgs scalars have mass $\sim m_A$)
\item two CP-odd scalars, linear combinations of the pseudoscalars 
      in the $\tilde W_3$ and $\tilde B$ $R$-partners: 
      we label these $A^0_{3}, A^0_4$.
\item one of the charged Higgs scalars, $H^{\pm}_3$
\item higgsinos and $R$-higgsinos
\end{itemize}

Five of these six states reside at the weak scale and are light only by comparison to the gauginos. For the (R-)higgsinos, charged Higgs, and one combination of CP-odd scalars ($A^0_4$), this weak-scale mass comes at tree level, $\sim \mu$ for the fermions and $\sim \lambda\,v$ for the scalars.   (Note that for the range of $\lambda$ we are interested in, this is high enough to avoid any direct bounds.) As we showed earlier, the Higgs boson can be made sufficiently heavy through the combination of a small $O(\lambda \mu/M)$ tree-level contribution and large radiative corrections. The mass of the remaining light state $A^0_3$, another CP-odd scalar, is insensitive to the Higgs vev and is instead set by a difference in soft masses. Because this state is independent of the Higgs vev, it does not  play a direct role in setting the strength of the phase transition, so its phenomenology may seem unrelated to the issues in this paper. However, the same combination of soft masses that enters into $m_{A^0_3}$ is also present in the mass of the $H^{\pm}_3$ and $A^0_4$ -- two states that play a large role in strengthening the phase transition. To increase the strength of the phase transition, $H^{\pm}, A^0_4$ should be as light as possible, which means we want to take the soft mass contribution to $m_{H^{\pm}_3}, m_{A^0_4}$ to be small. For this configuration of soft masses, $A^0_3$ will be much lighter than the weak scale (and possibly massless at tree level), so the viability of this parameter set given collider bounds is far from guaranteed.

The $A^0_3$ is a combination of gaugino $R$-partner fields, so it does not interact directly with SM fermions. The wino $R$-partner does have gauge interactions, while the bino $R$-partner does not. However, we are only interested in the neutral wino ($T_3 = 0$) $R$-partner component -- this component does not interact with $W^3_{\mu}$ at all and only has a quartic interaction with the charged $SU(2)$ gauge bosons. Furthermore, the mass eigenstate ($A^0_3$) coupling is suppressed by a mixing angle (squared) $\sim 1/2$ as only the wino $R$-partner component has this quartic interaction.

The remaining interactions are: a small trilinear $W_{\mu}^+W^{-\mu} (A^0_3)$ term, suppressed by $\sim v/M$, and Yukawa interactions with higgsinos and $R$-higgsinos. These interactions play no role in $A^0_3$ production, but they do permit $A^0_3$ to decay through loops. The only final state that can proceed through a higgsino loop is $A^0_3 \ra \gamma\gamma$. The gauge boson loop can lead to photons as well as light fermions, but the fermion component is subdominant due to the suppression by Yukawa couplings (and the overall coupling $\sim v/M$). For simplicity, we therefore assume $A^0_3\ra \gamma\gamma$ is the only available decay mode.

Given its interactions with standard model particles, it is quite challenging to produce the $A^0_3$. Because the $A^0_3$ does not interact with the $Z^0$, LEP places no constraint. The only production mechanism (at leading order in $v/M$) is vector boson fusion (VBF), $e^+e^- \ra \nu \bar{\nu} + W^+ W^- \ra \nu \bar{\nu} + (A^0_3)^2$, leading to the spectacular final state of missing energy plus four photons. However, even assuming a branching ratio to photons of 100\% for $A^0_3$, the rate is much too small to have been seen at LEP\@.

At hadron colliders, there are two ways to produce $A^0_3$; through VBF: $pp, p\bar p \ra jj + (A^0_3)^2 \ra jj + 4\gamma$, and from $s$-channel W production followed by the emission of a pair of $A^0_3$: $pp, p\bar p \ra W^{\pm} + (A^0_3)^2 \ra W^{\pm} + 4\gamma$. Provided some of the photons have high-$p_T$, such signals would be clearly visible above background.  The issue is whether the rate is high enough to generate more than a handful of events.

At the Tevatron, the rates (at leading order) for $ jj + (A^0_3)^2$ {\em before} VBF cuts are $\sim 0.5\,\fb$ for $m_{A^0,3} = 20\,\gev$, falling to $\sim 0.2\,\fb$ for $m_{A^0,3} = 50\,\gev$. At the LHC ($8\,\tev$), we find $\sigma(pp \ra  jj (A^0_3)^2)_{LO} \sim 26, 19\,\fb$ for $m_{A^0,3} = 20, 50\,\gev$ respectively. The significant increase in the rate at LHC is because $qq$ can initiate VBF production. After standard object identification and fiducial volume cuts (not even the usual VBF cuts) and accounting for realistic identification efficiencies (even if we look for fewer objects, like $\gamma\gamma+jj$), the Tevatron rates are too low to provide any bound. We arrive at the same conclusion for the $W^{\pm}+(A^0_3)^2$ process; the rate at the Tevatron, while slightly higher than the VBF process ($\sim 0.7\,\fb$ for $m_{A^0,3} = 20\,\gev$) is still too low to provide a meaningful bound given the Tevatron dataset and realistic object efficiencies. 

At the LHC, the VBF rate is high enough that a more thorough investigation is necessary. Di-jet plus multi-photon events would certainly fall under the scrutiny of the LHC Higgs searches. To test bounds on $A^0_3$ coming from Higgs diphoton limits, we generate signal events using the machinery of MadGraph4~\cite{Alwall:2007st}, and Delphes~\cite{Ovyn:2009tx}, then pass events through a mock CMS Higgs analysis. Though CMS and ATLAS have dedicated ``VBF" searches looking for 2 jets and 2 photons~\cite{cmsgammas, atlasgammas}, the analysis looking for a final state most similar to the $jj(A^0_3)^2$ final state, we find the VBF cuts imposed are too restrictive and hence the signal efficiency is extremely low $< 10^{-4}$. The more inclusive diphoton Higgs searches have looser cuts, but we also find them to be not particularly sensitive to our signal, $\epsilon \sim 10^{-3}$. The lack of sensitivity is due to a few reasons: While there are more photons in our signal, the photons themselves have lower energies and sit in a more crowded environment as opposed to the diphoton signal ($\sim 125\,\gev$ Higgs) the cuts were designed for. Hence the leading photon $p_T$ cut, photon-jet isolation, and photon-photon isolation requirements remove much more signal compared to a SM Higgs. A heavier $A^0_3$ would pass the cuts more efficiently, but has a smaller production cross section.

While we find no firm bounds on the $A^0_3$ from Higgs (or other) searches, the LHC rate is certainly large enough that a dedicated multi-photon plus jets (or plus $W^{\pm}$) may well be worthwhile.

\section{Discussion}

We have seen that electroweak baryogenesis can be achieved in 
the minimal supersymmetric $R$-symmetric model with:
\begin{itemize}
\item an electroweak phase transition strength $\phi_c/T_c \gsim 1$
\item Higgs mass $\simeq 125$~GeV
\item induced baryon asymmetry $\eta_{10} \gtrsim 1$.
\end{itemize}
The central ingredients are the new superpotential couplings,
Eq.~(\ref{eq:extrayuk}), where we required 
$\lambda_W \simeq -\lambda_B \simeq 2$ to achieve a strong enough 
first order phase transition simultaneous with $\mh \simeq 125$~GeV\@.

That we needed modestly large $\lambda$s providing substantial
trilinear interactions between the Higgs boson and the additional
scalars in $R_u$, $R_d$, $\Phi_B$ and $\Phi_W$ is perhaps not
particularly surprising, given the degree of freedom counting given
in Ref.~\cite{Carena:2004ha}.  Larger $\lambda$ couplings are potentially
problematic if the theory is run to higher scales, though this is
beyond the scope of this paper.  However, there are a few comments we can 
make on this point.  Interestingly, the interactions between the 
chiral adjoints $\Phi^a$ and the Higgs/$R$ superfields are also present 
in models with $\mathcal{N} = 2$ supersymmetry involving 
$\mathcal{N} = 2$ vector supermultiplets interacting with
$\mathcal{N} = 2$ hypermultiplets (e.g.~\cite{Polonsky:2000zt,Fox:2002bu}).
There the coupling strength is determined by $\mathcal{N} = 2$ 
supersymmetry to be $\sqrt{2} g$ for the appropriate gauge group
(times $Y_i^2$ for $U(1)_Y$).   
This is somewhat smaller than, but not that far from the superpotential
coupling strengths of interest in our case.  If we had taken the 
$\mathcal{N} = 2$ limit for the superpotential couplings,
they would evolve identically to the gauge couplings up to 
the explicit $\mathcal{N} = 2$ breaking terms.  This suggests
that the renormalization group evolution is not necessarily as
drastic as, say, the superpotential coupling for $S H_u H_d$ 
in the NMSSM.  It would be interesting to investigate this further,
and to determine the role of the relative \emph{signs} of these 
couplings on the evolution.

\section*{Acknowledgements}

We thank A.~Nelson and S.~Su for discussions. 
GDK and AM thank the Aspen Center of Physics where part of this
work was completed.  RF was partially supported by funding from 
NSERC of Canada.
GDK were supported in part by the US Department of Energy 
under contract number DE-FG02-96ER40969 and by NSF 
under contract PHY-0918108.
AM is supported by Fermilab operated by Fermi Research Alliance, 
LLC under contract number DE-AC02-07CH11359 with the 
US Department of Energy. 
YT was supported in part by the NSF through grant PHY-0757868.

\appendix
%\begin{appendix}

\onecolumngrid
%\begin{widetext}

\section{Scalar potential}
\label{sec:appa}

The contributions to the scalar potential comes from the superpotential, the D-term, supersoft terms and the scalar soft masses. The superpotential is
\begin{align}
W_{RSSM} &=  (H^+_u\,R^-_u)\Big( \mu_u + \lambda_{uB} A_0  - \frac{\lambda_{uW}}
{2} A_3 \Big) - (H^0_u\, R^0_u)\Big( \mu_u + \lambda_{uB} A_0  + \frac{\lambda_{
uW}}{2} A_3 \Big) \nonumber \\
&~~~~~ + (R^+_d\,H^-_d)\Big( \mu_d + \lambda_{dB} A_0  - \frac{\lambda_{dW}}{2} 
A_3 \Big) - (R^0_d\, H^0_u)\Big( \mu_d + \lambda_{dB} A_0  + \frac{\lambda_{dW}}
{2} A_3 \Big) \nonumber \\
& ~~~~~+ \frac{A^-}{\sqrt 2}\Big( \lambda_{uW}\,(H^+_u\,R^0_u) + \lambda_{dW}\,(
R^+_d\,H^0_d) \Big) - \frac{A^+}{\sqrt 2}\Big( \lambda_{uW}\,(H^0_u\,R^-_u) + \lambda_{dW}\,(R^0_d\,H^-_d) \Big) + c.c. 
\end{align}

From the superpotential, we get the usual potential $V \supset -\sum_i |F_i|^2$,
 where $i$ runs over all of the above superfields.\\

The D-term contribution to the Lagrangian is
\begin{eqnarray}
\textrm{D-term} &=& -\frac{g'^2}{8}(\hupsq + \huosq - \Rumsq - \Ruosq - \hdosq - \hdmsq + \Rdpsq + \Rdosq)^2 \nonumber \\
& & - \frac{g^2}{8} \Bigg| \hupsq - \huosq + \Ruosq - \Rumsq + \hdosq - \hdmsq + \Rdpsq - \Rdosq - 2 i (A_1^* A_2 - A_2^* A_1)\Bigg|^2 \nonumber \\
& & -\frac{g^2}{8} \Bigg| \hupc\huo + \Ruoc \Rum + \hdmc \hdo + \Rdpc \Rdo - 2 i A_2^* A_3 + c.c \Bigg|^2 \nonumber \\
& & -\frac{g^2}{8} \Bigg| i \huoc \hup + i \Rumc \Ruo + i \hdmc \hdo + i \Rdoc \Rdp - 2 i A_3^* A_1 + c.c. \Bigg|^2
\end{eqnarray}
The supersoft terms are
\begin{eqnarray}
\textrm{supersoft}  &=& - M_1^2 (A_0 + A_0^*)^2 - \sqrt{2}g' M_1 (A_0 + A_0^*)\frac{1}{2}(\hupsq +\huosq-\hdosq - \hdmsq - \Rumsq-\Ruosq + \Rdpsq + \Rdosq)\nonumber \\
& &  - M_2^2 (A_j + A_j^*)^2 - \sqrt{2} g M_2 (A_3 + A_3^*)\frac{1}{2}(\hupsq -\huosq +\hdosq - \hdmsq - \Rumsq +\Ruosq + \Rdpsq - \Rdosq)\nonumber \\
& & - \sqrt{2} g M_2 (A_1 + A_1^*)\frac{1}{2}(\hupc \huo + \hdoc \hdm + \Ruoc \Rum + \Rdpc  \Rdo + c.c.)\nonumber \\
& & - \sqrt{2} g M_2 (A_2 + A_2^*)\frac{1}{2}(-i \hupc \huo -i \hdoc \hdm -i \Ruoc \Rum -i \Rdpc  \Rdo + c.c.)\nonumber \\
\end{eqnarray}

We further decompose the neutral fields $A_0$ and $A_3$, CP even and CP-odd pieces
\begin{equation}
A_0 = \frac{s_0 +i\,p_0}{\sqrt 2}, \quad A_3 = \frac{s_3 + i\,p_3}{\sqrt 2}.
\end{equation}
The CP even pieces will mix with the CP-even Higgs scalars $h_u, h_d$, while the CP-odd pieces only mix among themselves. \\

The soft mass terms are the usual MSSM soft masses, plus equivalent $R_u, R_d$ soft masses. The only tricky soft masses are for the $A_i$
\begin{align}
V_{soft} \supset m^2_{s0} (A_0\,A_0^*)  + \frac{m^2_{p0}}{2} (A^2_0 + A^{*,2}_0)  + m^2_{st}(A_i A^*_i) + \frac{m^2_{pt}}{2} (A^2_i + A^{*,2}_i),
\label{eq:softmm}
\end{align}
where $i$ is an $SU(2)$ index. In the language of Sec.~\ref{sec:cpv}, $m^2_{p0}, m^2_{pt}$ are holomorphic soft masses (and can potentially carry a phase), while $m^2_{s0}, m^2_{st}$ are non-holomorphic, and therefore purely real, soft masses. 

\section{Light field potential}

Starting with the full potential given in the previous appendix, we first integrate out the scalars with mass $M$, keeping terms of $O(1/M_D)$ and $O(1/M_D^2)$. The Higgs scalar soft masses $m^2_{H_u}$, $m^2_{H_d}$ can be removed by enforcing electroweak symmetry breaking. Specifically, under the set of assumed relations between various parameters laid out in Eq.~(\ref{eq:numass}), we find. 
\begin{align}
m^2_{H_u} + \mu^2 &= m^2_A\,\cos^2{\beta} - \frac{\mu\,M_Z\,v\,\sin^2{\beta}(\cos{\theta}\,\lambda_W - 2\sin{\theta}\lambda_B)}{2\,\sqrt 2 M}  + O\Big(\frac 1 {M^2} \Big) \nn \\
m^2_{H_d} + \mu^2 &=  m^2_A\,\cos^2{\beta} - \frac{\mu\,M_Z\,v\,\cos^2{\beta}(\cos{\theta}\,\lambda_W - 2\sin{\theta}\lambda_B)}{2\,\sqrt 2 M}  + O\Big(\frac 1 {M^2} \Big). \nn \\
\end{align}
Though we have only shown the modifications to $O(1/M)$, we retain terms to $O(1/M^2)$ when calculating the tree-level potential. Focusing on neutral, CP even Higgs fields and plugging in the expressions for $m^2_{H_u}, m^2_{H_d}$, we find the tree level potential to be
\begin{eqnarray}
V_{tree} & = & \frac{m^2_A}{2}\phi^2\sin^2{(\beta-\chi)} + \frac{\mu\,M_Z\,\phi^2( (\phi^2-v^2)\cos{2\chi} - v^2\cos{2\beta} )}{8\sqrt 2 M v} (2\lambda_B\sin{\theta} - \lambda_W\cos{\theta}) \nn \\
&  &{} + \frac{\phi^2\,v^2}{1024 M^2} \Big(2\,M^2_Z\,\cos{2\beta} \cos{2\theta}(\cos{2\beta} + 2\cos{2\chi})(4\lambda^2_B - \lambda^2_W + 4\lambda_B\lambda_W\,\tan{2\theta}) \nn \\ 
& & \qquad\qquad - (4\lambda^2_B + \lambda^2_W)(M^2_Z\,(1 + 4\cos{2\beta}\cos{2\chi} + \cos{4\beta}) -32\,\mu^2) \nn \\
& & \qquad\qquad \qquad\qquad - 64 M^2_Z\,\cos{2\beta}\cos{2\chi}\,(\sin^2{\theta}(m^2_{p0} + m^2_{s0}) + \cos^2{\theta}(m^2_{st} + m^2_{pt} ) )  \Big) \nn \\
& & {} + \phi^4\frac{(2M^2_Z\,\cos^2{2\chi}\,(\sin^2{\theta}(m^2_{p0} + m^2_{s0}) + \cos^2{\theta}(m^2_{pt} + m^2_{st})) - \mu^2\,v^2\,(4\lambda^2_B + \lambda^2_W) ) }{64\,M^2 v^2}  \nn \\
& & {} +  \phi^6 \frac{M^2_Z \cos^2{2\chi}(2\lambda_B \sin{\theta} - \lambda_W\,\cos{\theta})^2}{256\,M^2\,v^2}, 
\label{eq:lighttree}
\end{eqnarray}
where $\phi_u, \phi_d$ are the up- and down-type Higgs fields and $\phi^2 \equiv \phi^2_u + \phi^2_d$,\,$\tan{\chi} \equiv \phi_u/\phi_d$

In addition to the tree-level potential, our calculation also requires the field dependent masses for all of the light states (mass $\ll M$) in the theory.  By field-dependent we mean that the Higgs fields $\phi_u$ and $\phi_d$ are {\em not} set to the zero-temperature vacuum values $v_u, v_d$. The mass matrices and eigenstates are given in the subsequent subsections. In all expressions we neglect any $O(1/M)$ or smaller pieces. This truncation is justified because the effects of these states on the Higgs potential are already suppressed by loop factors.

\subsection*{CP-odd, Charge Neutral Higgs Scalars}

The field-dependent mass matrix for the four $R=0$, neutral, CP-odd fields $(a_u, a_d, p_0, p_3)$ is block diagonal and is given below:
\begin{eqnarray}
\left(\begin{array}{cccc}
m^2_A\,\cos^2{\beta} & m^2_A\, \cos{\beta}\sin{\beta} & 0 & 0 \\
m^2_A\, \cos{\beta}\sin{\beta} & m^2_A\,\sin^2{\beta}  & 0 & 0 \\ 
0 & 0 & m^2_{s0}-m^2_{p0} + \frac{\lambda_B^2}{2}(\phi^2_u + \phi^2_d) & \frac{ \lambda_B\lambda_W}{4}(\phi^2_u + \phi^2_d) \\
0 & 0 & \frac{\lambda_B\lambda_W }{4}(\phi^2_u + \phi^2_d) &  m^2_{st}-m^2_{pt} + \frac{\lambda_W^2}{8}(\phi^2_u + \phi^2_d) \\
\end{array}\right), \nonumber \\
\end{eqnarray}

The pseudoscalar Higgs mass matrix has a zero eigenvalue corresponding to the Goldstone boson eaten by the $Z$. The three massive eigenvalues are
\begin{align}
m^2_{A^0,2} &= m^2_A \nn \\
m^2_{A^0,3} & = \Delta_A + \frac{\lambda_B^2 + \frac{1}{4} \lambda_W^2}{4}(\phi^2_u + \phi^2_d) -  \sqrt{\Delta_A'^2 +  \Delta_A' (\lambda_B^2 - \frac 1 4 \lambda_W^2)(\phi^2_u + \phi^2_d) + (\lambda_B^2 + \frac 1 4 \lambda_W^2)(\phi^2_u + \phi^2_d)^2}, \nn \\
 m^2_{A^0,4} & =  \Delta_A + \frac{\lambda_B^2 + \frac{1}{4} \lambda_W^2}{4}(\phi^2_u + \phi^2_d) +  \sqrt{\Delta_A'^2 +  \Delta_A' (\lambda_B^2 - \frac 1 4 \lambda_W^2)(\phi^2_u + \phi^2_d) + (\lambda_B^2 + \frac 1 4 \lambda_W^2)(\phi^2_u + \phi^2_d)^2}, \nn \\
 & ~~\Delta_A = m^2_{s0} + m^2_{st} - m^2_{p0} - m^2_{pt}, \nn \\
 & ~~\Delta'_A = m^2_{s0} - m^2_{st} - m^2_{p0} + m^2_{pt}, 
\end{align}
up to corrections of order $1/M$.

\subsection*{Charged Higgs Scalars ($R=0$)}

For the charged Higgs scalars, we started with four states that all mix with each other: $H_u^+, H_d^{-*}, A^+, A^{*-}$ + c.c. One combination of $A^+, A^{*-}$ is heavy and gets integrated out. The remaining three-by-three mass matrix is

\begin{eqnarray}
\left(\begin{array}{ccc}
m^2_A\,\cos{\beta}^2  &  m^2_A\,\cos{\beta}\sin{\beta}  & 0 \\
 m^2_A\,\cos{\beta}\sin{\beta}  & m^2_A\,\sin{\beta}^2 & 0 \\
0  & 0 & m^2_{st} - m^2_{pt} + \frac{\lambda^2_W}{8}(\phi^2_u + \phi^2_d) \\
 \end{array}\right),\nn \\
\end{eqnarray}

This mass matrix has a zero eigenvalue corresponding to a Goldstone degree of freedom. The remaining eigenvalues are
\begin{align}
m^2_{H_{\pm, 2}} & = m^2_A\nn \\
m^2_{H_{\pm, 3}} & =  m^2_{st} - m^2_{pt} + \frac{\lambda_W^2}{8}(\phi^2_u + \phi^2_d) 
\end{align}

\subsection*{Neutral $R=2$ scalars}

The mass matrix for the complex, neutral scalar 
$R$-partners of the Higgs fields is:
\begin{eqnarray}
\left(\begin{array}{cc} 
\mu^2 + m^2_{R} + \frac{\lambda_B^2 + \frac 1 4 \lambda_W^2}{2}\, \phi^2_u &\frac{1}{2} \lambda_B\lambda_W \,\phi_u\,\phi_d \\
\frac{1}{2} \lambda_B\lambda_W \,\phi_u\,\phi_d & \mu^2 + m^2_{R} + \frac{\lambda_B^2 + \frac 1 4 \lambda_W^2}{2} \phi^2_d
\end{array}\right), \nonumber
\end{eqnarray}
The eigenvalues become particularly simple if we take $m_{Ru} = m_{Rd} = m_R$
\begin{align}
m^2_R + \mu^2,\ m^2_R + \mu^2 + \frac{\lambda_B^2 + \frac{1}{4} \lambda^2_W}{2}(\phi^2_u + \phi^2_d), 
\end{align}
in which case only one of the states has a $\phi$-dependent mass.

\subsection*{Charged $R=2$ scalars}
The charged $R$-scalars do not mix with any other states, so they have a simple mass term

\begin{align}
(\mu^2 + m^2_{R} + \frac{\lambda_W^2}{4}\,\phi^2_u)\,|R^-_u|^2 + (\mu^2 + m^2_{R} + \frac{\lambda_W^2}{4}\,\phi^2_d)\,|R^+_d|^2
\end{align}

\subsection*{CP even neutral scalars}
Though we start with four CP-even, charge neutral scalars ($h_u, h_d,s_0, s_3)$, two have mass $\sim M$ and are integrated out of the low-energy theory. The remaining two states to form the light Higgs boson $h$ and its heavier cousin $H$. The heavy Higgs boson has mass $\sim m_A$ and plays no role in determining the strength of the phase transition. The light Higgs boson is massless at lowest order at $O(M^0)$, but is lifted to nonzero mass (at tree level) by $O(1/M, 1/M^2)$ terms in $V_{tree}$ (Eq.~(\ref{eq:lighttree}) ). The full expression is long and not very insightful.  The mass matrix simplifies significantly at large $\tan{\beta}$, leading to the expression given in Eq.~(\ref{eq:mhlite}).

\section{Ino mass matrices}

The neutralino mass matrix starts as
\begin{eqnarray}
 \left(\begin{array}{cccc}
                M_2 & 0 & -\frac{g\, \phi_u}{2} & \frac{g\, \phi_d}{2} \nonumber \\
                0 & M_1 & \frac{g'\, \phi_u}{2} & -\frac{g' \,\phi_d}{2} \nonumber \\
                \frac{\luW \phi_u}{2\sqrt{2}} & \frac{\luB \phi_u}{\sqrt{2}} & -\muu & 0 \nonumber \\
                \frac{\ldW \phi_d}{2\sqrt{2}} & \frac{\ldB \phi_d}{\sqrt{2}} & 0 & -\mud \\
\end{array}\right), \nonumber\\
\end{eqnarray}
Under our parameter assumptions, once the mass $M$ scalars and gauginos are integrated out, the matrix collapses to a diagonal two-by-two matrix with entries $\pm \mu + O(1/M)$.\\

The chargino mass matrix combining $\tilde W^+, \tilde R^+_d$ with $\tilde A^-, \tilde H^-_d$ is originally
\begin{align}
\label{eq:cha1}
& \left(\begin{array}{cc}
                M_2 & \frac{g\, \phi_d}{\sqrt{2}} \nonumber \\
                -\frac{\ldW \phi_d}{2}& \mud \\
\end{array}\right), \end{align}
After our usual parameter assumptions and upon integrating out the heavy states, we are left with a single entry
\begin{align}
m^2_{\chi^{\pm}_1} = \Big|\mu + \frac{M_Z}{8\,M\,v}\Big(2\lambda_B\sin{\theta}(\phi^2_d - \phi^2_u) + \cos{\theta}((1+4\sqrt 2)\phi^2_d - \phi^2_u) ) \Big)\Big|^2,
\end{align}
where we have retained the $O(1/M)$ piece within  $m^2_{\chi^{\pm}_1} $ because it will be relevant for the CP-violation section. The extra $2\sqrt 2\phi^2_d$ in the above expression comes from integrating out the gauginos (at tree level), in addition to the heavy scalars.

Similarly, the matrix combining $\tilde A^+, \tilde H^+_u$ with $\tilde W^-, \tilde R^-_u$ begins as:
\begin{align}
\label{eq:cha2}
&\left( \begin{array}{cc}
        M_2 & \frac{\luW \phi_u}{2} \\
        \frac{g\, \phi_u}{\sqrt{2}} & \muu \\
\end{array}\right), 
\end{align}
becoming
\begin{align}
m^2_{\chi^{\pm}_2} = \Big|\mu + \frac{M_Z}{8\,M\,v}\Big(2\lambda_B\sin{\theta}(\phi^2_d - \phi^2_u) + \cos{\theta}(\phi^2_d - (1 + 4\sqrt 2)\phi^2_u) )\Big) \Big|^2,
\end{align}
once the heavy states are removed.

\twocolumngrid

\end{document}